\documentclass[12pt,aps,prd,preprint,showkeys]{revtex4}
\usepackage[utf8x]{inputenc}
\usepackage{amsmath}
\usepackage{amsfonts}
\usepackage{amssymb}
\usepackage{graphicx}
\usepackage{csvsimple}
\usepackage{multirow} 
\usepackage[toc,page]{appendix}

\usepackage{layouts}

\begin{document}
	
	\title{Machine Learning model driven prediction of the initial geometry in Heavy-Ion Collision experiments.}
	\author{ Abhisek Saha$^1$}
	\author{Debasis Dan$^2$}
	\author{Soma Sanyal$^1$}
	
	\affiliation{$^1$ School of Physics, University of Hyderabad, Gachibowli, Hyderabad, India 500046}

	\affiliation{$^2$ Microsoft India (R\& D) Pvt. Ltd. Microsoft Campus, ISB Road, Gachibowli, Hyderabad, India 500032}

	\begin{abstract}
		
		
		
We demonstrate high prediction accuracy of three important properties that
 determine the initial geometry of the  heavy-ion collision (HIC) experiments by 
 using supervised Machine Learning (ML) methods. These properties are the impact 
 parameter, the eccentricity and the participant eccentricity. Though ML 
 techniques have been used previously to determine the impact parameter of these 
 collisions, we study multiple ML algorithms, their error spectrum, and sampling 
 methods using exhaustive parameter scans and ablation studies to determine a combination of 
 efficient algorithm and tuned training set that gives multi-fold 
 improvement in accuracy for all three different 
 heavy-ion collision models. The three models chosen are a transport model, a 
 hydrodynamic model and a hybrid model. The motivation of using three different 
 heavy-ion collision models was to show that even if the model is trained using a 
 transport model, it gives accurate results for a hydrodynamic model as well as a 
 hybrid model. We show that the accuracy of the impact parameter prediction 
 depends on the centrality of the collision. With the standard application of ML training methods,  
 prediction accuracy is considerably low for central collisions. Our method increases this accuracy by multiple folds. 
 We also show that the eccentricity prediction accuracy 
 can be improved by inclusion of the impact parameter as a feature in all these 
 algorithms. We discuss how the errors can be minimized and the accuracy can be 
 improved to a great extent in all the ranges of impact parameter and 
 eccentricity predictions.

		
	\end{abstract}

	\maketitle

	\section{Introduction}
	
	Ever since the first run of the heavy-ion collision experiments, a lot of studies have been carried out describing and analyzing the data that we get from these experiments \cite{expt1, expt2,expt3}. 
	 The beam energy scan program of RHIC at BNL(Brookhaven National Laboratory) runs the collider experiments using Au-Au nuclei at collision energies from $7.7$ GeV to $200$ GeV \cite{qgp, BES1,BES2,BES3}. One of their important aims is to search for the critical point in the QCD phase diagram \cite{phase}. The matter created in these experiments has a high baryon density. On the other hand, Pb-Pb collisions are conducted at LHC at collision energies $2.76$ TeV \cite{276alice}, $5.02$ TeV \cite{502alice}, the aim of these experiments is to examine the high-temperature region of the QCD phase diagram.
	  The distribution of particles in the initial stage is different for different collision systems and this affects the final stage particle spectra and the anisotropic flows \cite{epsv}. The primary data we get from these experiments are the transverse momentum($p_{T}$) spectra, the rapidity($y$) spectra, the pseudorapidity($\eta$) spectra, the particle-antiparticle ratios and the multiplicity fluctuations. Some phenomena e.g., the anisotropic flows can be obtained directly from these data. But some parameters are difficult to calculate directly from the experimental data. These are the impact parameter, the initial state geometry parameters, (e.g., eccentricities) event plane angles, etc. The impact parameter is the distance between the centers of the colliding nuclei on the transverse plane of the collision. In experiments, the data is always studied with respect to the centrality of the collision as we get different spectra for different centrality collisions. The high centrality collisions are those where the impact parameter is close to zero i.e., the head-on collisions. The peripheral collisions refer to the higher values of the impact parameter. The collision centrality plays an important role in determining the final particle spectra. The multiplicity distribution of different species is observed to be dependent on the centrality of the collision. In ref. \cite{BES2,BES_cen}, the multiplicity fluctuations at different centralities are studied at RHIC energies and in ref. \cite{276alice, 502alice,544alice}, the same has been studied at collision energies $2.76$ TeV, $5.02$ TeV, and $5.44$ TeV respectively. The centrality is not a property that can be attained directly from the experiments, but it can be calculated with the help of theoretical modeling by using the Glauber model\cite{Glauber} or some other similar model. The impact parameter as well as the initial geometry of the collision is difficult to determine experimentally. This is true especially for the more central collisions. That is why there are various proposals to determine the impact parameter. Apart from different simulations and algorithms, neural networks have also been proposed to determine the impact parameter from the experimental data \cite{david}.  
	
The determination of the impact parameter is related to the charged multiplicity produced during the heavy ion collision. The charged particle multiplicity has a contribution from hard and soft collision processes which in turn depend on the number of participants and also on the number of binary collisions. If $x$ is the fraction of contribution from hard processes then the charged particle multiplicity per unit pseudorapidity can be expressed as,
	\begin{equation}
		\frac{dN_{ch}}{d\eta} = n_{pp}\left[  (1-x)\frac{N_{part}}{2} + xN_{coll}  \right]
		\label{eq:1}
	\end{equation}
here $n_{pp}$ is the multiplicity per unit rapidity in pp collisions and $N_{coll}$ is the number of binary NN collision. The number of participant nuclei $N_{part}$ can be expressed as a function of the impact parameter \cite{Glauber, branges}. If $T_{A}(s)$ is the thickness function of nucleus $A$ i.e., the probability density function of finding nucleons in $A$, then the number of participants in $A$ at the transverse position $s$ can be found out by multiplying with the probability of binary nuclei-nuclei collision with the nucleons of the nucleus B at the same position $(b-s)$ where $b$ is the impact parameter. So, the total number of participants can be expressed as,
	\begin{align}
		N_{part}(b) =& \int T_{A}(s) (1-exp[-\sigma_{inel}^{NN}T_{B}(b-s)])ds \nonumber\\ & + \int T_{B}(b-s) (1-exp[-\sigma_{inel}^{NN}T_{A}(b)])ds 
		\label{eq:2}		
	\end{align}
	Here the total number of participating nuclei is found out by summing over the contribution from nucleus A and nucleus B. Using Eq.\ref{eq:1} and Eq.\ref{eq:2}, the impact parameter hence centrality can be estimated by fitting the multiplicity spectra. In this method, the multiplicity fitting must be done for every event to obtain its centrality. An easier way of getting the centrality is to use machine learning models. Machine learning has been invoked to determine the impact parameter from the experimental data in several papers\cite{ML1, ML2, ML3}. Using machine learning, we can automate the whole process and the impact parameter can be calculated in an efficient way. The advantage of using machine learning (ML) is that it requires less computational power as well as computational time. So, this makes the process more agile. Most of the work in this field is related to the deep neural network algorithms. The convoluted neural network has also been used to make predictions about the impact parameter \cite{pang}. The first paper to demonstrate the importance of neural network analysis for improving the accuracy of the determination of the impact parameter is ref. \cite{bass}. Using an ANN (Artificial Neural Network) or CNN (Convolutional Neural Network) they have  effectively  determined the impact parameter, but these networks require the tuning of hundreds of parameters. This makes the process computaionally expensive. On the other hand, different non-neuronal ML models like SVM, RandomForest, kNN etc require lesser parameters to produce results with similar accuracy as the ANN or CNN models. Thereafter many papers have explored various Machine Learning algorithms to obtain more accurate results for the impact parameter.  
	
	In this study, we have analyzed various machine learning algorithms (ML) and provided a rigorous comparison on the accuracy and efficiency of these algorithms using well defined techniques of machine learning to show a critical gap 
	in their prediction accuracy for central collisions. 
	We have mainly focused on three properties, impact parameter, eccentricity, and participant eccentricity. We  analyze the errors in the predictions and discuss the causes that lead to these errors. 
	We find that the accuracy is less for the low impact parameters. This is an already known problem in the determination of the impact parameter. We provide a custom sampling method that shows significant 
	improvement in accuracy over commonly used sampling methods in the ML community. We have also used a particular HIC model for training, while the data from two different HIC models have been used to make the predictions.
	This indicates that for a well-defined training data set, the predictions for the impact parameter using the ML model are model independent.

	   In this study, the transverse momentum($p_{T}$) spectra are taken as features and the impact parameter, eccentricity, and participant eccentricity are taken as the target variable which the model must predict. 
	   We have used the AMPT (A Multi-Phase Transport) model to generate the transverse momentum spectra of Au-Au collision events at $200 GeV$ collision energy \cite{ampt}. 
	   The charged particle multiplicity has been studied previously using the AMPT model\cite{ampt_multi}. As the target variables are known for fitting, we use supervised machine learning algorithms. 
	   Also, the target is a continuous variable so it can have any real value, hence we  use the regression algorithms.

 The focus of our study would be predicting the impact parameter and the eccentricity. Eccentricity is one of the parameters which gives us the initial geometrical shape of the collision region. This also affects the elliptic flow of produced particles which is one of the important observables used to study collective behavior in heavy-ion collisions. In ref. \cite{ecc2v2}, the effects of eccentricity fluctuation on the elliptic flow is studied at $\sqrt{s}=200$ GeV for Au-Au and Cu-Cu collisions. In a recent study\cite{amptecc}, the flow-harmonics are studied as a function of different components of initial anisotropy using the AMPT model. In our study, the learnings and experiences gathered by the ML models from the impact parameter prediction, is passed onto predict the eccentricity of the initial stage of the heavy-ion collision system. We have also looked at how the inclusion of the impact parameter as a feature, affects the prediction accuracy. We have made predictions of the initial state anisotropy, using the impact parameter. The initial state anisotropy is given by\cite{branges},
	\begin{equation}
		\epsilon_{n}(b) = \frac{<r^{n}cos(n\phi -n\psi)>}{r^{n}}
		\label{eq:eps}
	\end{equation}
	here $r=\sqrt{x^{2}+y^{2}}$, $\psi=tan^{-1}\frac{y}{x}$, n=2 gives the eccentricity and n=3 gives the triangularity.
	The above eccentricities are with respect to the reaction plane. We have also trained the model to predict participant plane eccentricity which is given by\cite{ecc2v2},
	\begin{equation}
		\epsilon_{part} = \frac{ \sqrt{\sigma_{y}^{2} - \sigma_{x}^{2}+4\sigma_{xy}^{2}} } {\sigma_{y}^{2} + \sigma_{x}^{2}}
	\end{equation}
	here $\sigma$'s are the variances of the positions of the particles,
	$\sigma_{x}^{2} = <x^{2}>-<x>^{2}$, $\sigma_{y}^{2} = <y^{2}>-<y>^{2}$ and $\sigma_{xy} = <xy>-<x><y>$. Here $<..>$ is the average over the transverse plane.
	
The AMPT is a transport model which has been used extensively to model the different stages of the heavy ion collision from the initial collision dynamics to the final stage hadron dynamics. However, like all models, it has certain drawbacks. There are alternate simulations based on hydrodynamics which also give reliable outputs which match well with the data. In this study, we have taken the results from other models too. This is to test if the predictions of the ML algorithms depend crucially upon the nature of the model used. Our results show that as long as the models reflect the experimental data accurately, the ML algorithms do not distinguish between the different models. An ML algorithm trained on a specific model gives pretty accurate results when tested with the data generated by a different model.  	
	
The two other heavy ion collision models used in this study are VISH2+1(Viscous Israel Stewart Hydrodynamics (2+1) dimension) \cite{vish1} and a hybrid model made of a hydro evolution model and a hadronic cascade model \cite{iEBE}. These two models are different from the AMPT model which is used to train the ML algorithms. So, the ML models train from the $p_{T}$ spectra and the impact parameter data of AMPT events and predicts impact parameters using test data of $p_{T}$ spectra from the VISH2+1 and the hybrid model. We choose the initial conditions of different models such that they generate the $p_{T}$ spectra close to the one that is obtained in the actual experiments. In this way, we are examining the efficiency of the ML algorithms in a model-independent manner. However, the model-independency is limited only to those models which generate the $p_{T}$ spectra close to the experimental $p_{T}$ spectrum. The $p_{T}$ spectra of the hydro and the hybrid model are fitted with the experimental $p_{T}$ spectra to measure the effectiveness of ML models on the experimental data.
	
	In section II, we give a brief description of the heavy-ion collision models used in this study. In section III, we talk about the ML models used in this study. We also describe the parameters used to check the accuracy of different ML models. The learning process of various algorithms as well as the tuning of the hyperparameters are given in this section. We have also used rebalancing techniques to improve the accuracy of the results. These rebalancing techniques are discussed in this section. 
	 Section IV discusses the results and the predictions made by the ML models of the eccentricity and the participant eccentricity. It also discusses the ranges of eccentricity where optimum accuracy has been observed. The efficiency of predicting the impact parameter using unknown data of different HIC models and experimental data is discussed in this section. In the end, we show how the accuracy can be improved by rebalancing the dataset. We then summarize the paper in section V.

	
	\section{Event Generation} 
	\subsection{The AMPT model}
The	AMPT is a publicly available heavy-ion collision model which generates heavy-ion collision events. It is often used to understand the results obtained from experiments and it has successfully given the results which match well with the experimental observations \cite{ampt}. There are two versions of AMPT. In both, the initial condition is generated by the HIJING model \cite{hijing1,hijing2,hijing3}. Here the initial configuration of nucleons is determined by the Glauber model with a Wood-Saxon nuclear distribution. Particle production from two colliding nuclei is given in terms of two processes. In hard processes, the momentum transfer is larger and they are described by pQCD and they produce minijets. The soft processes are those where the momentum transfer is lower and described by the non-perturbative process by the formation of strings. Of the two models, in the default version, the partons recombine with their parent strings after the end of interaction in the partonic state and forms hadrons using the Lund String fragmentation model \cite{lund}. In the string melting (SM) version of AMPT, the strings are converted to their valence quarks and antiquarks. The partonic stage interactions are described by Zhang's Parton Cascade (ZPC) where the interactions are described by the Boltzmann equations \cite{zpc}. The scattering cross-section of the parton interactions is calculated using pQCD. The simplified relation between total parton elastic scattering cross-section and the medium induced screening mass is taken as,
	\begin{equation}
		\sigma \approx \frac{9\pi \alpha_{s}^{2}}{2\mu^{2}}
	\end{equation}
	where $\alpha_{s}$, the strong coupling constant and $\mu$, the screening mass, taken as 0.33 and $3fm^{-1}$ respectively for a total cross-section of $3$ mb. When the partons stop interacting, they are hadronized by a quark coalescence model. Here the nearest quark-antiquark pair is converted into a meson and the three nearest quarks or antiquarks are converted into a baryon or an antibaryon.
	The hadronic dynamics are described by the ART (A Relativistic Transport) model \cite{ART}. We have used both the versions of the AMPT model. In all the collision setup, we use Au-Au collision at $200$ GeV collision energy. Different centralities are considered for different purposes. The other settings are the same as the parameters taken in these references \cite{ampt_multi,ampt_settings}.
	
	\subsection{The VISH2+1 model}
	VISH2+1 is a publicly available code where the evolution of the system created in heavy-ion collisions is described by relativistic causal viscous hydrodynamics \cite{viscous, vish1,vish2}. The code has been tested extensively and it has successfully reproduced the results from experiments \cite{vish_expt}. The initial distribution is taken from the Glauber model in terms of energy-momentum tensor $T^{mn}$. Then it solves the local energy-momentum conservation equation $d_{m}T^{mn}=0$ where,
	\begin{equation}
		T^{mn}=eu^{m}u^{n}-p\Delta^{mn}+\pi^{mn} 
	\end{equation}
	Here $\Delta^{mn}=g^{mn}-u^{m}u^{n}$, $u^{m}$ and $u^{n}$ are the velocity components, and $p$ is the pressure.
	$\pi^{mn}$ is the viscous shear pressure which follows the evolution equation,
	\begin{equation}
		D\pi^{mn}=\frac{1}{\tau_{\pi}}(2\eta\sigma^{mn}-\pi^{mn})-(u^{m}\pi^{nk} - u^{n}\pi^{mk})Du_{k}
	\end{equation}
	$D= u^{m}d_{m}$ and the symmetric and traceless shear tensor is given by, $\sigma^{mn}= \frac{1}{2}(\nabla^{m}u^{n}+ \nabla^{n}u^{m})-\frac{1}{3}\Delta^{mn}d_{k}u^{k}$. The pressure $p$ and the energy density $e$ are related by the equation of state(EoS), which is used to solve the hydrodynamic equations. There are three different EoS used in this study, EoS-L, SM-EOS Q, and s95p-PCE. The EOS L is based on lattice QCD data where a smooth crossover transition connects the QGP state to the chemically equilibrated hadron resonance gas(HRG) state\cite{vish_expt}. The SM-EOS Q is the smoothed version of the EOS Q where a first-order phase transition with a vacuum energy (bag constant) connects the non-interacting QGP state to the chemically equilibrated HRG state \cite{eosq}. The s95p-PCE equation of state is obtained from fits to lattice QCD data for crossover transition at high temperatures and to a partial chemical equilibrium system of the hadrons at low temperatures\cite{s95p}.

	In the Israel-Stewart \cite{IS1,IS2} framework, the generalized hydrodynamic equation of an energy-momentum tensor $T^{mn}$, together with viscous pressure contributions $\pi^{mn}$ is solved with a collision time scale $\tau_{\pi}$(relaxation time).
	The longitudinal boost-invariance is implemented and seven equations are solved, $3$ for the $T^{\tau\tau}$, $T^{\tau x}$ and $T^{\tau y}$ and $4$ for the $\pi^{mn}$'s. Here a flux-corrected transport(FCT) algorithm called SHASTA (Sharp And Smooth Transport Algorithm) \cite{SHASTA} is used to solve the hydrodynamic and kinetic equations. It has two stages. In the transport stage, the multidimensional calculations are simplified in terms of geometric interpretation which is followed by an anti-diffusive or corrective stage. This technique is also applied in codes like AZHYDRO \cite{azhydro}. The final spectra are obtained on a freeze-out hypersurface where the fluid stops interacting. The freeze-out is computed using the Cooper-Frye procedure \cite{cooper} at a decoupling temperature $T_{dec}$. Here iSpectra (iS) code is used, which is a fast Cooper-Frye particle momentum distribution technique that gives discrete momentum distribution of the desired hadron species \cite{iSS}. We get events of emitted hadrons similar to the events generated in experiments which are then used for ML model predictions.
	
	\subsection{Hybrid model}
	We have used the iEBE-VISHNU code package \cite{iEBE} which is a hybrid model made by combining a (2+1)-dimensional viscous hydrodynamic model and a hadronic cascade model. Instead of using the whole package, we have used the modules separately for better handling of inputs and outputs. The output of the iS particle sampler obtained at the end of hydro evolution, is used for hadronic re-scatterings. Here, the UrQMD after-burner package is used to serve this purpose. After the particalization, the hadrons are produced on the hypersurface with individual production time and location. The position and momenta along with the ids are then written in a standard OSCAR1997A format which is suitable for hadronic re-scattering\cite{oscar}. This is done using the Oscar to UrQMD converter routine. This also propagates all the hadrons backward in time so that all of them have a fixed initial time and the Boltzmann collision integral can be performed in the UrQMD model. 
	
	Ultra-Relativistic Quantum Molecular Dynamics (UrQMD) is a transport model where the dynamics of the hadrons are modeled \cite{urqmd1, urqmd2}.  UrQMD can generate a whole collision system starting from the nuclear collision to the hadronic spectra but here we have only used it to get the hadronic evolution. The interaction among the hadrons is evaluated using the Boltzmann equation for the distribution of all hadrons. The system evolves through binary collisions or by 2-N-body decays. $53$ baryon species and $24$ different meson species, along with their resonances, antiparticle states and isospin-projected states are considered in the UrQMD interactions. The interaction among the hadrons and their resonances in this model are described in reference \cite{urqmd2}.
	
	

\section{Machine Learning Methods}
\subsection{ML Algorithms and Tuning of Hyperparameters}
As mentioned in the introduction, there are various ML algorithms that we have tested for this study e.g. k-NearestNeighbors, Gradient Boosting Regression, Decision Trees etc. . Details of these ML algorithms are available in ref. \cite{mlalgorithms}. The accuracy of these models has been tested using standard measures such as R-square, the Root Mean Square Error (RMSE), the Mean Squared Error (MSE), and the Mean Absolute Error (MAE). After running various ML algorithms, we find that although all the algorithms give very good predictions for the impact parameter, only three of them perform well for the eccentricity prediction. Hence, we concentrate only on these three algorithms. They are the k-NearestNeighbors(kNN), ExtraTrees Regressor(ET) , and the Random Forest Regressor(RF) model. In kNN model, the target is predicted by doing an local interpolation of the target associated with the k nearest neighbors of the training dataset\cite{knn1}.  ET and RF are the kinds of ensemble method. In RF, the decision trees are made during the training and a mean of the ensemble is calculated\cite{RF1}. In ET,  randomized decision trees are considered that are made of sub-samples of the training dataset\cite{etr}. We have used a 10-fold cross-validation (CV) to obtain a robust estimate of the parameters\cite{cv}. This also gives a bias-variance trade off.  

Since we have used these ML algorithms for studying the data from three different HIC models, we have standardized the data before processing them. In this study, we are using the $p_{T}$ spectra of charged particles as features in the dataset. The $p_{T}$ spectra is obtained for the mid-rapidity region with a rapidity window of $-0.5$ to $0.5$. All the $p_{T}$ bins have a  different range of values. The difference is more significant when we compare a lower $p_{T}$ bin with a higher $p_{T}$ bin. Thus, it is important to make them standardized.  This makes the model compatible with a new dataset coming from a different HIC model. Two types of scaling are used in this study, i) the Standard Scaler or Z-score normalization and ii) the Min-Max Scaler\cite{normalize}. 
	To use both these scaling techniques, we have used python sklearn.preprocessing library \cite{sklearn}. In most of the cases discussed in this study, we observe that the Z-score method provides us an accuracy greater than the min-max scaling by $4-6\%$. So, in all the cases, we have used the Z-score standardization. 
	
After standardization, the $p_{T}$ spectra serve as the features in the dataset and the target variables are the impact parameter, the eccentricity, and the participant eccentricity. When the impact parameter is used as the target variable, only the $p_{T}$ spectra is used as feature variables. For the other targets, the predicted impact parameter is included in the dataset as a feature variable as all the other targets have a dependency on the impact parameter. In this way, all the dependent variables can be measured just by giving the $p_{T}$ spectra as inputs. A standard training and test set separation was done for model evaluation.

\begin{figure}
	\includegraphics[width=0.6\linewidth]
	{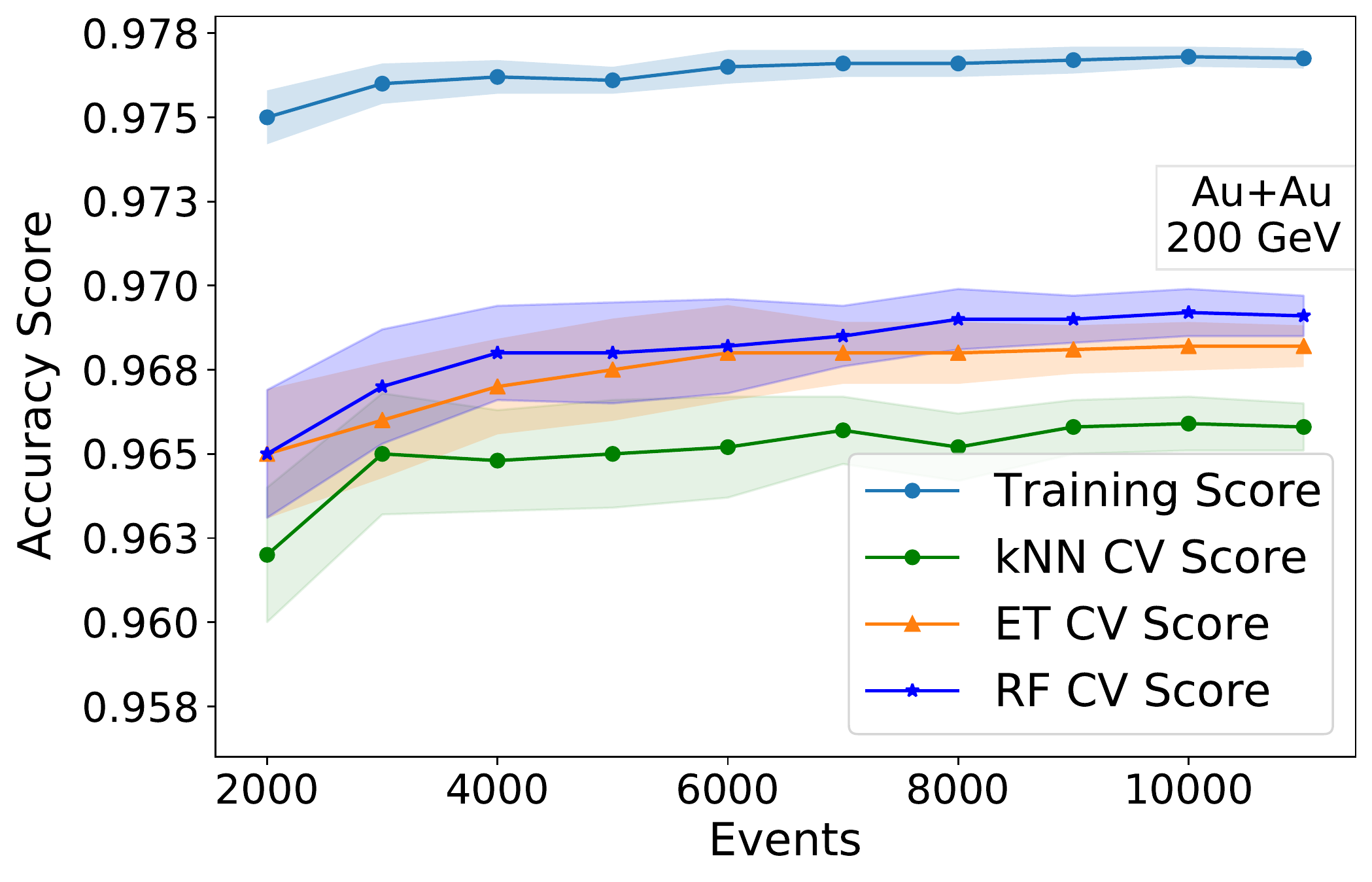}
	\caption{The learning curve of kNN(green dots), ET (orange triangles), and RF (blue stars) model. Accuracy score as a function of the number of events are shown which attains saturation after $3000$ events for the training set (sky line), and after $6000-8000$ events for the test sets. The shaded region is the standard deviations}
	\label{fig:learning}
\end{figure}
	
It is important to have enough events to achieve the best accuracy without consuming too much computing resources. 
The learning curve of a machine learning model tells us how effectively a model is learning throughout its running time. We present the learning experience as a function of events. In Fig.\ref{fig:learning}, the learning curves of a kNN (green circles), ET (orange triangles), and RF (blue stars) model are shown where the changes in the Cross-Validation (CV) accuracy are represented with the number of event iterations. The training score curve is shown only for the kNN model (sky color circles) which shows the accuracy while fitting the training data to the model. In the training case, the accuracy comes to a saturation very early around $3000$ events. While in the cases of test data accuracy shown by the other curves saturates around $6000$ to $8000$ events. All the learning in this study are performed over $10k$ events. The shaded area represents the standard deviations in the accuracy score.

\begin{figure}
		\includegraphics[width=0.5\linewidth] {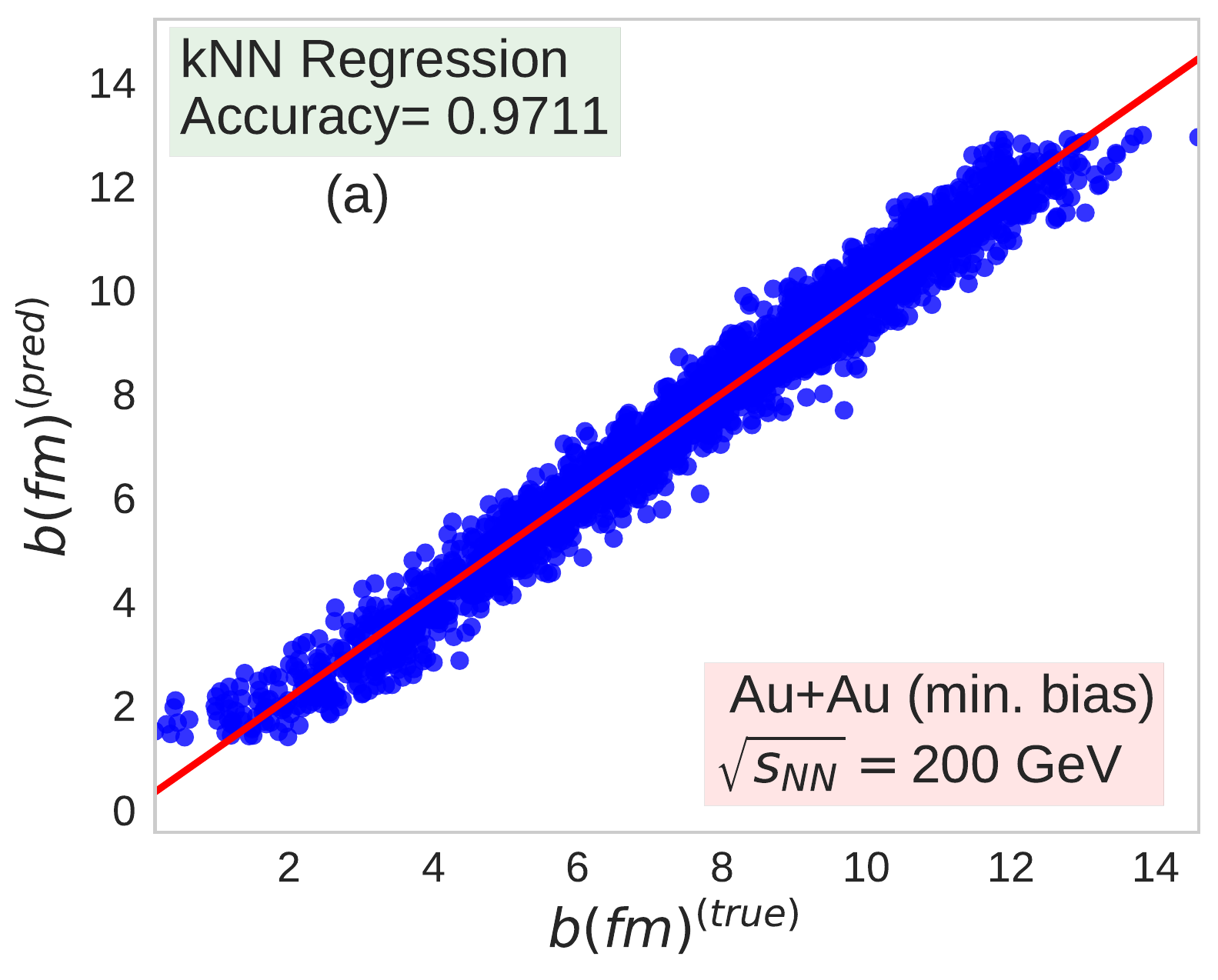}\includegraphics[width=0.5\linewidth] {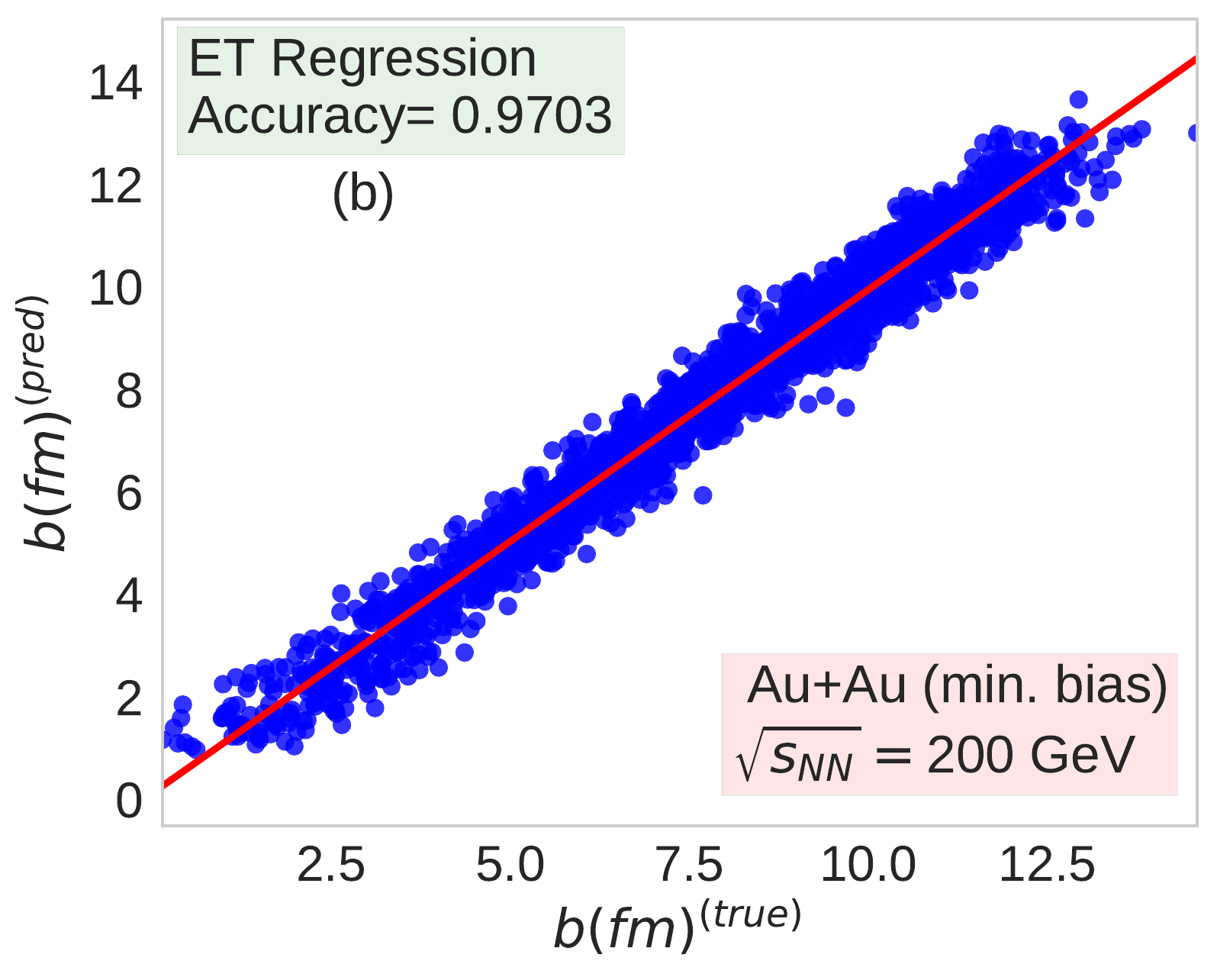}
		
		\includegraphics[width=0.5\linewidth] {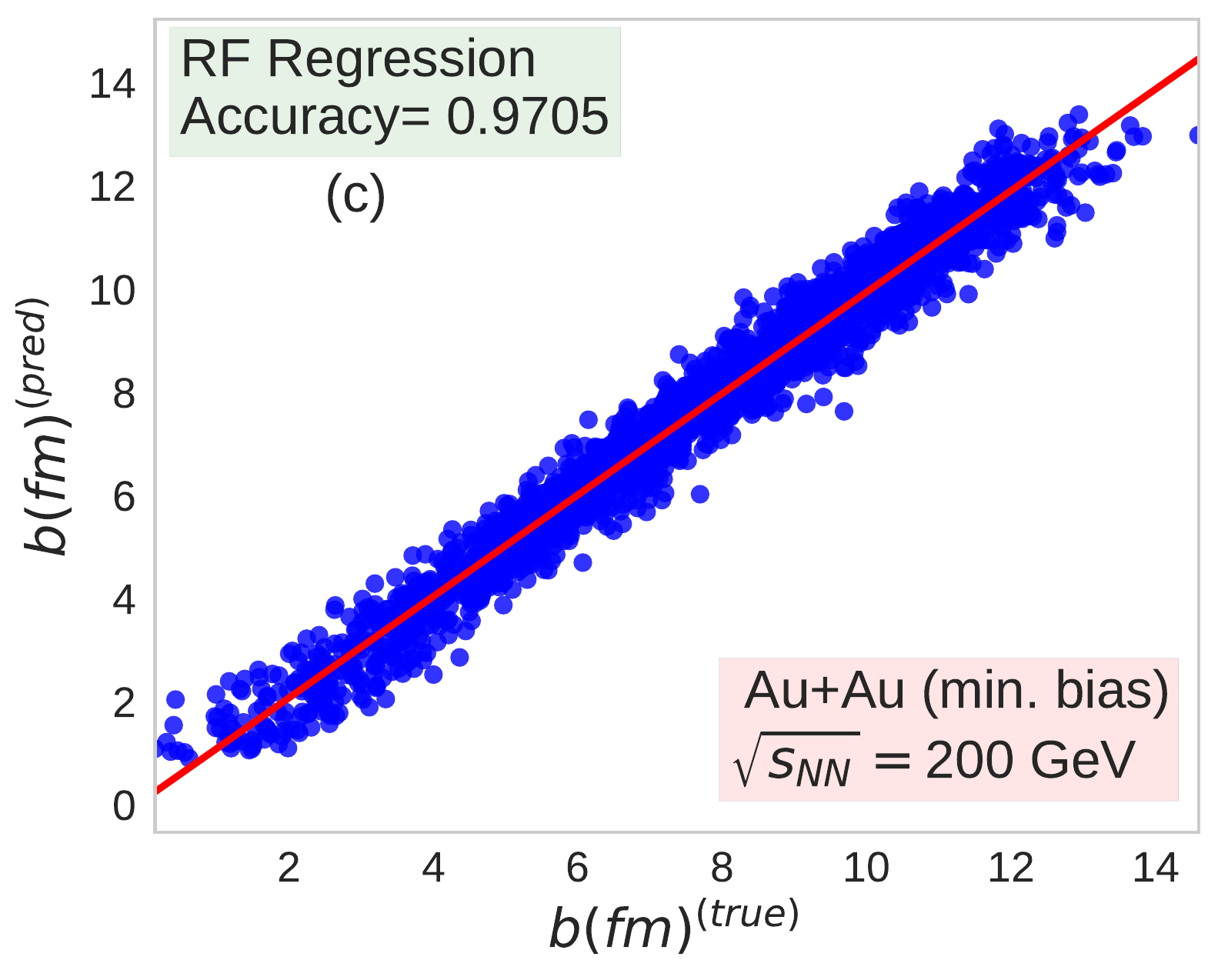}\includegraphics[width=0.5\linewidth] {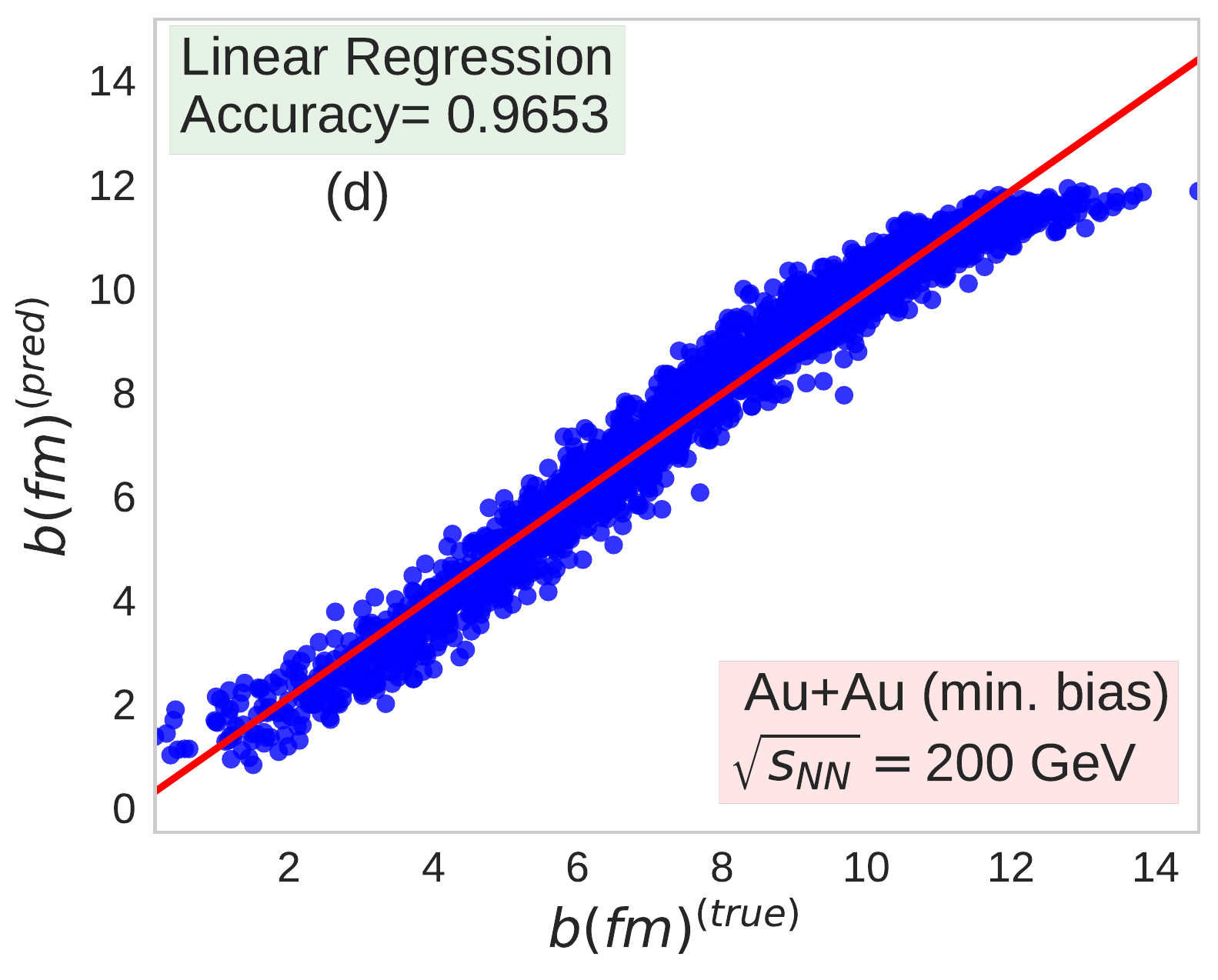}
		\caption{Impact parameter prediction using kNN(a), ET(b), RF(c) and LR(d) model with their accuracy score $97.11\%$, $97.03\%$, $97.05\%$ and $96.53\%$ for events of Au+Au system at collision energy $200$ GeV. These plots are obtained for a random train and test set split of input events.}
		\vspace{-10pt}
		\label{fig:b_error}
	\end{figure}

In Fig.\ref{fig:b_error}, the accuracy plots for impact parameter predictions using kNN(a), ET(b), RF(c), and Linear Regression(LR)(d) models are shown. The ML models are trained using the charged particle $p_{T}$ spectra data of AMPT-SM model. Linear Regression algorithm finds a linear relationship between a dependent and one or more independent variable\cite{lr}. The prediction is performed using a test dataset containing $p_{T}$ spectra of more than $4000$ events of minimum bias Au+Au collision at $200$ GeV.  The red line drawn here is the optimum accuracy line and the blue points are the predictions made by the model.  The accuracies achieved are $97.11\%$, $97.03\%$, $97.05\%$ and $96.53\%$ for kNN, ET, RF, and LR model respectively. All of these accuracies are observed for a random train-test dataset split. The 10-fold cross-validation scores of these models are $97.04\%$, $97\%$, $97.01\%$, and $96.56\%$ respectively. We get accuracy of more than 95\% for the kNN, ET and RF models when the ML models are trained using the default AMPT model data, .  
 In the case of impact parameter predictions, most of the machine learning algorithms give a fair level of accuracy without tuning any of the hyperparameters except in certain critical impact parameter regimes. 



\begin{figure}
		\includegraphics[width=.6\linewidth]{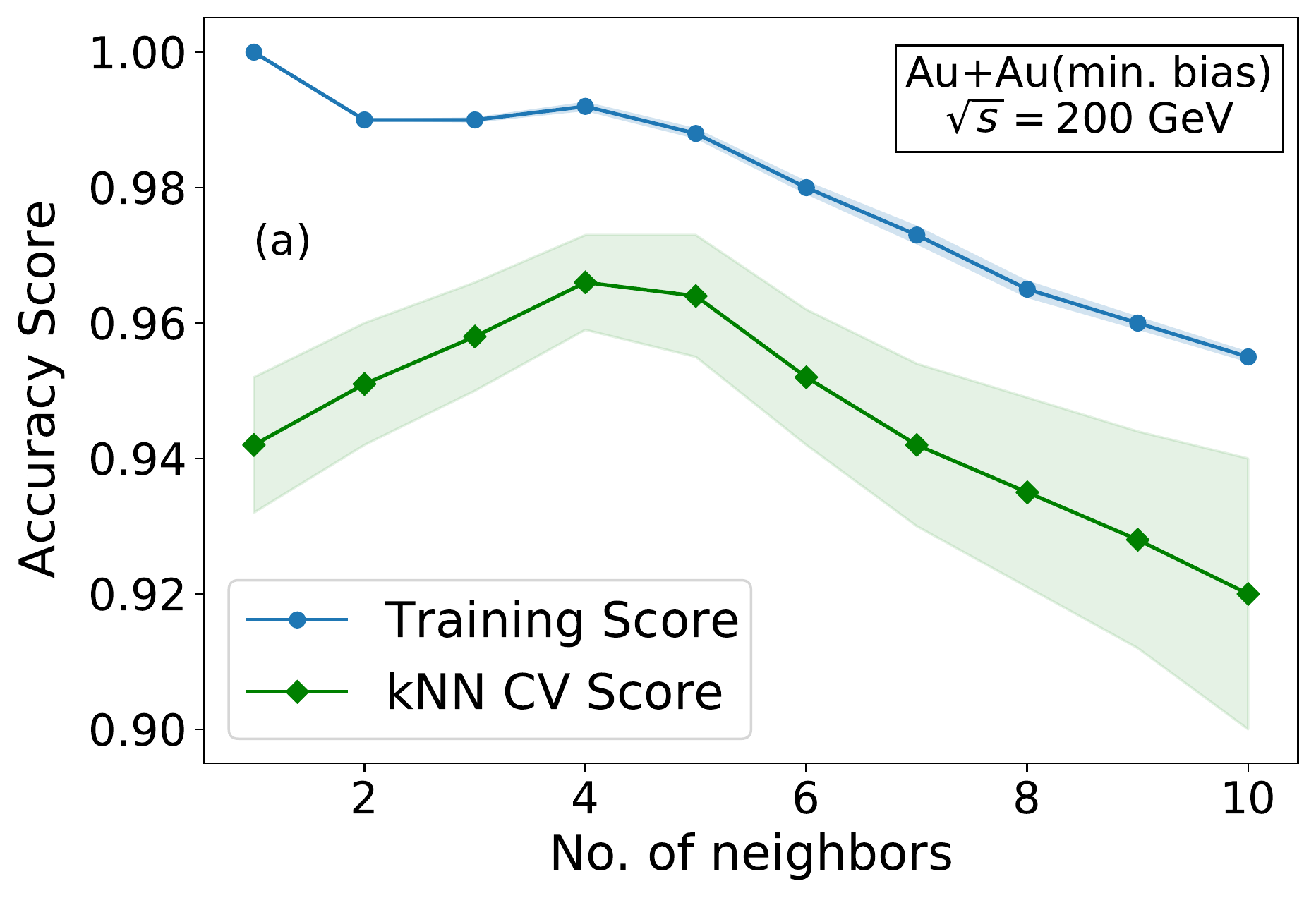}
		\includegraphics[width=.6\linewidth]{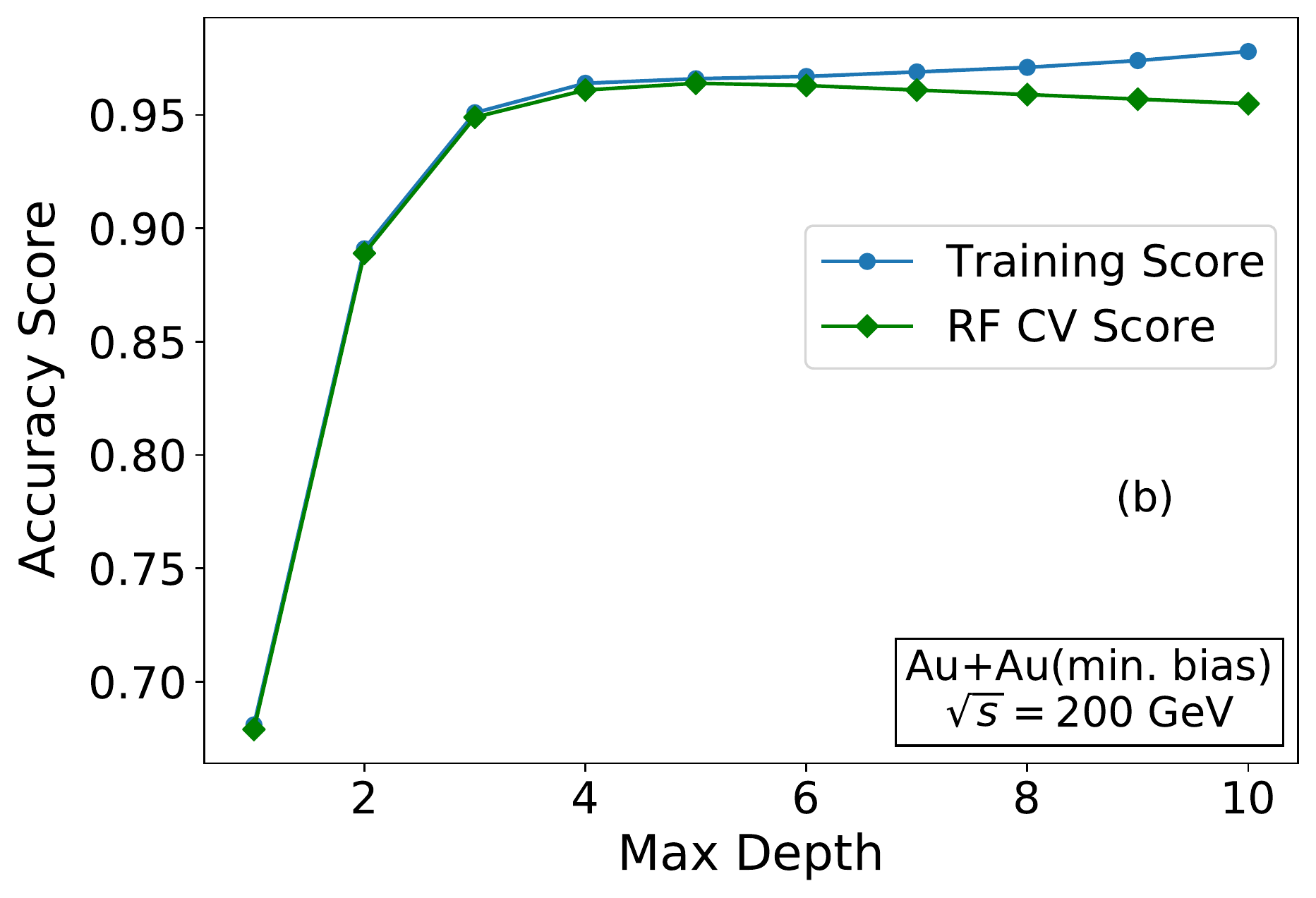}
		\caption{Change in accuracy as a function of hyperparameters. a) kNN model with the number of nearest neighbors hyperparameter, b)Random Forest with max depth hyperparameter}
		\label{param}
	\end{figure}

	It is known that the choice of parameters can affect the accuracy of a model. Hyperparameter tuning was done to fix the parameters with minimum error validation set. In Fig. \ref{param}(a), the change in the accuracy of a kNN model is shown as a function of the number of nearest neighbors hyperparameter. For every configuration, the model is trained using $12000$ events of minimum bias Au+Au collision, and the impact parameter is taken as the target variable. The highest accuracy is attained by the model when the number of nearest neighbors is $4$ or $5$. This is shown by the green curve which gives the 10-fold cross-validation score and the shaded region is the standard deviation. The training score shown by the blue line has a score of $1.00$ when the number of nearest neighbors is $1$. This is a case of overfitting. For the random forest (RF) model (see Fig. \ref{param}(b)), the choice of hyperparameter is the maximum number of levels of the tree. We find that the accuracy saturates for the hyperparameter value of $4$ or $5$. Like the RF model, we get the maximum CV score of the ET model when the max-depth hyperparameter is $4$ or $5$.
	Although the above-mentioned parameters are the ones that hamper the accuracy most, we fix the other hyperparameters by  running the RandomSearchCV function of the sklearn library and checking the accuracy for a different combination of hyperparameters.

\begin{figure}
		\includegraphics[width=0.8\linewidth] {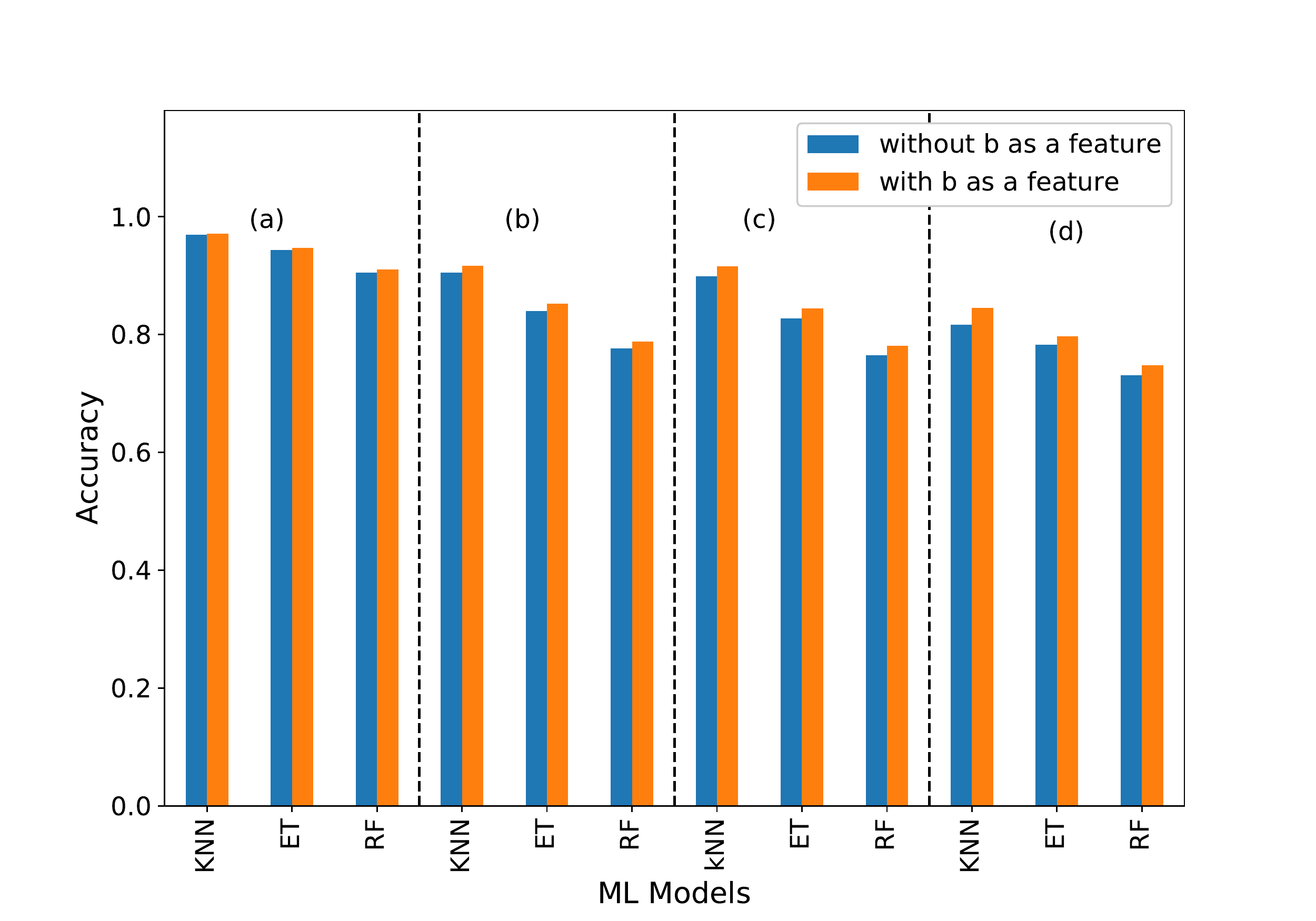}
		\caption{Effect on the eccentricity prediction accuracy by the inclusion of impact parameter as a feature for different centrality(\%), a) 0-10\%, b) 10-40\%, c) 40-80\%, d) Min. bias events. The orange bar represents accuracy with impact parameter as a feature and blue bars represent accuracy without impact parameter as a feature. }
		\label{fig:ecc_b1}
	\end{figure}

	As discussed earlier, in Fig.\ref{fig:ecc_b1}, we see how the inclusion of the impact parameter as a feature affects the accuracy of eccentricity prediction. As is seen in earlier studies that eccentricity depends on the centrality of the collision. Here we have found that by the inclusion of the impact parameter as a feature, the accuracy increased in all the centrality ranges.  

\begin{figure}
		\includegraphics[width=0.6\linewidth]{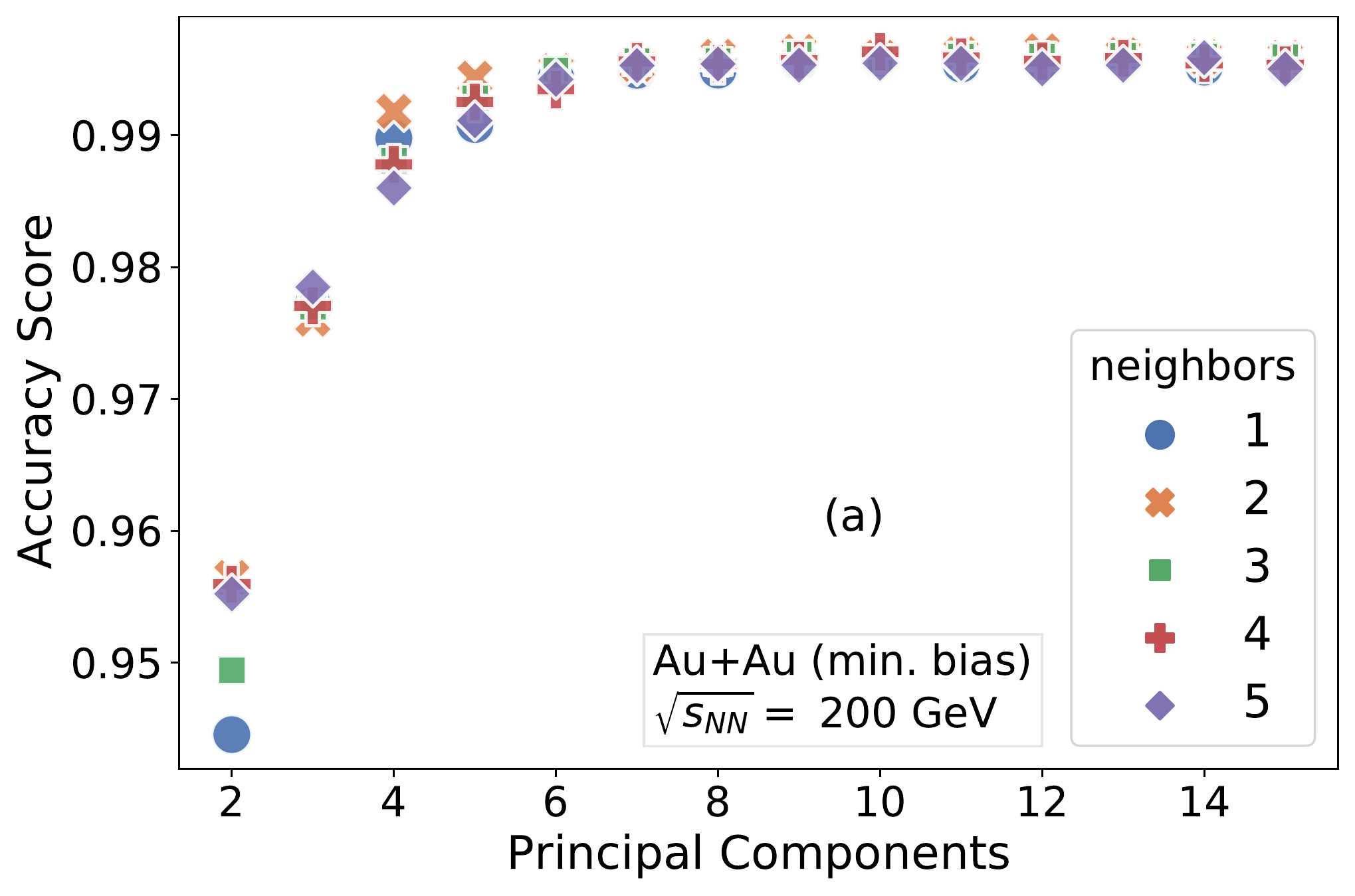}
		\includegraphics[width=0.6\linewidth]{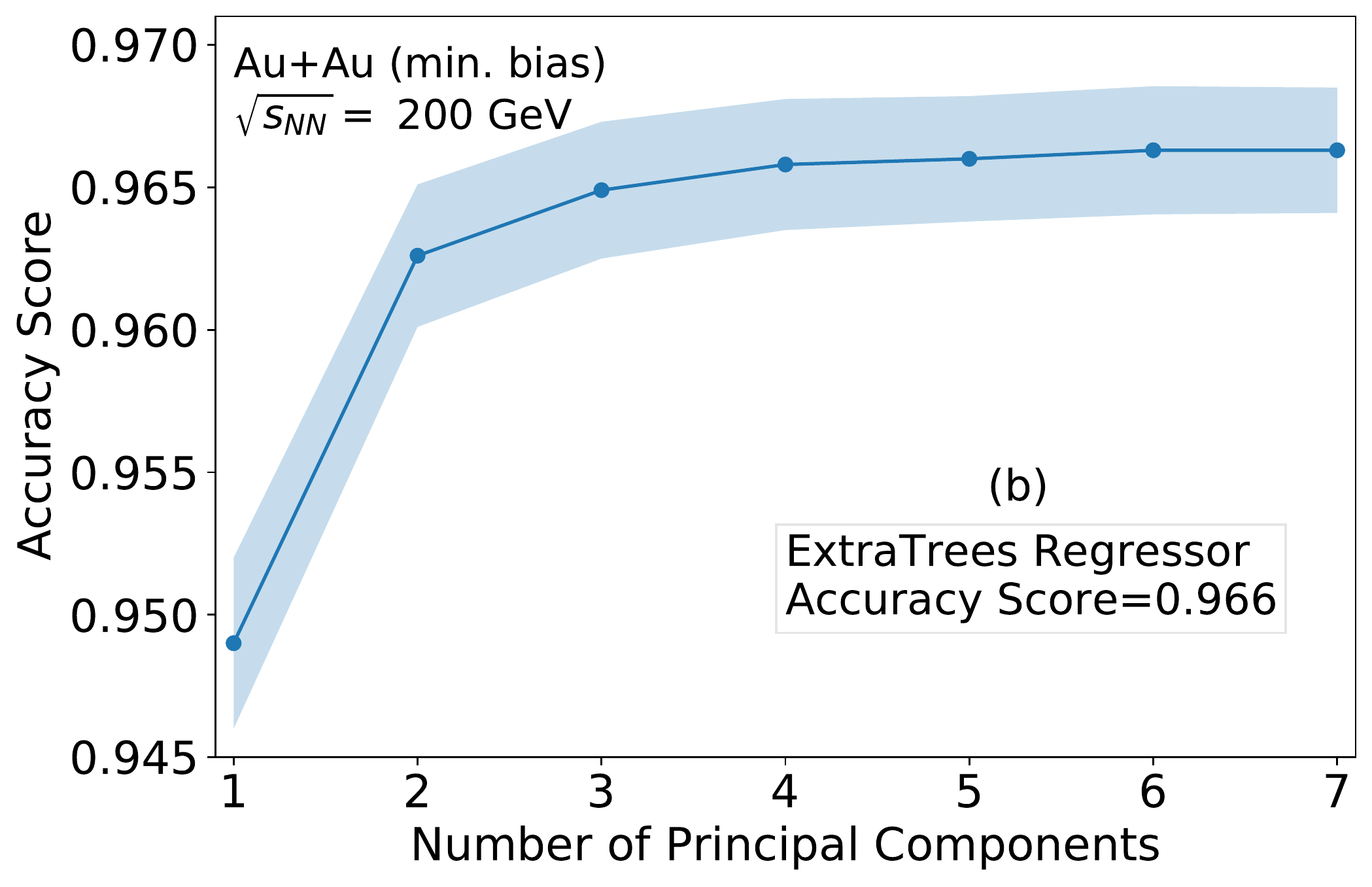}
		\caption{a) Accuracy of a a) kNN model and b) an ET model as a function of the number of principal components used }
		\label{fig:pca}
	\end{figure}

	The errors in ML can be reduced by determining the highly correlated features in the data. The Principal Component Analysis (PCA) is the most popular technique used for feature reduction of a large dataset\cite{Cadima}. In this work, we have tried the 'SelectFromFeature' function from the sklearn library and PCA method to reduce the colinearity and compared the outcomes to the already achieved accuracy using all the features. We have only shown the result of PCA method. In Fig. \ref{fig:pca}(a), the accuracy score of a kNN model, and in Fig.\ref{fig:pca}(b), the accuracy score of an ET model is shown as a function of the number of principal components is used. Here the accuracy is observed for the impact parameter predictions using $p_{T}$ distribution dataset of $12000$ minimum bias Au-Au collision events at $200$ GeV collision energy. The saturation in the accuracy score is achieved for the use of $7$ or more principal components in both cases.	Also, using 7 components, a variance coverage of $95\%$ can be achieved in the case of impact parameter predictions.  So, it is safe to use $7-8$ principal components to get a good amount of accuracy without losing any major information. We used 7-8 components for the impact parameter determination. We also found that at least $10$ features or $10$ principal components are needed to obtain an accurate result for the eccentricity and the participant eccentricity. This is expected as the data that we are using is the transverse momentum data. Since the impact parameter is known to be correlated with the transverse momentum data hence, we need a smaller number of features to obtain a high accuracy of prediction \cite{grebieszkow} for the impact parameter as compared to the eccentricity. In all the eccentricity predictions we will use PCA function to transform the features.

	\subsection{Custom resampling for unbalanced training set}
	The $p_{T}$ spectra we used as a feature are comprised of imbalanced datasets. As we have considered the $p_{T}$ spectra of minimum bias events, there are a smaller number of events for lower impact parameter values. Thus, the event distribution of $p_{T}$ spectra is left-skewed. The imbalance in the data affects the prediction accuracy of the impact parameter and eccentricity in the lower b region ($b\le 1 fm$). As is well known in the literature, the impact parameter is not directly accessible to the experiments. Bass et. al \cite{bassip} have pointed out that though most of the experimental observables are dependent on the impact parameter, the different methods of impact parameter estimation are usually optimized for the larger impact parameter range. This means that the expermental results for head-on collisions pertaining to the lower impact parameter range will have higher errors due to the inaccuracy of the impact parameter calculations. Currently, efforts are being made to improve the prediction of the impact parameter in the lower impact parameter range. This is very important as there are considerable experimental results from head-on collisions which can be better analysed with an improved prediction of the impact parameter in the lower range. So our aim is to improve the accuracy of the impact parameter in the lower range by balancing the data set appropriately.

	There are a few sampling techniques in machine learning for rebalancing datasets e.g., SmoteR, ADASYN \cite{smote, adasyn}. These are python packages that increase(over-sampling) or decrease(under-sampling) the minority and majority data class respectively using the neighboring data. We evaluated both techniques with all possible hyperparameter combinations. The results discussed in the next section (Section IV C) indicate that we do not have a sufficient increase in the accuracy of the predictions and there is a high chance of central events being predicted as non-central ones. 
	
	We then adopt a method of rebalancing the data set using class weights, where different classes are the different impact parameter regimes. The various combinations of distribution region and weights were evaluated through an exhaustive grid search. Based on test set minimum error, we selected events with impact parameter $<= 1.0 $ fm to be in category $1$ and the rest in category $2$. The weights assigned to the two classes are in the ratio $4:1$. This technique has helped us to reduce the errors further and the results are discussed in detail in the next section (subsection C).

	
	\section{Results and Discussions}
\subsection{Impact parameter and eccentricity prediction}	
	
	\begin{figure}
		\includegraphics[width=0.6\linewidth] {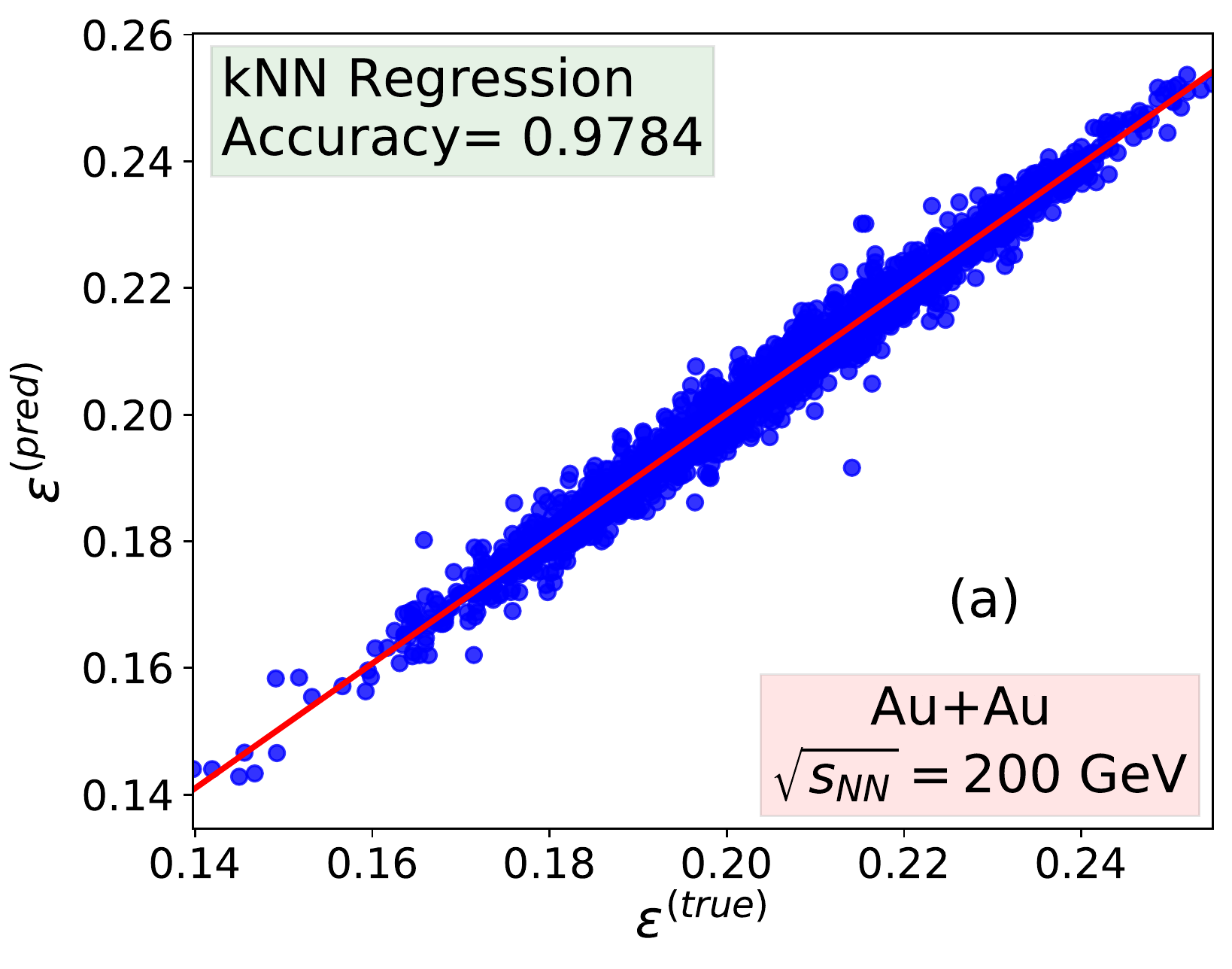}
		\includegraphics[width=0.6\linewidth] {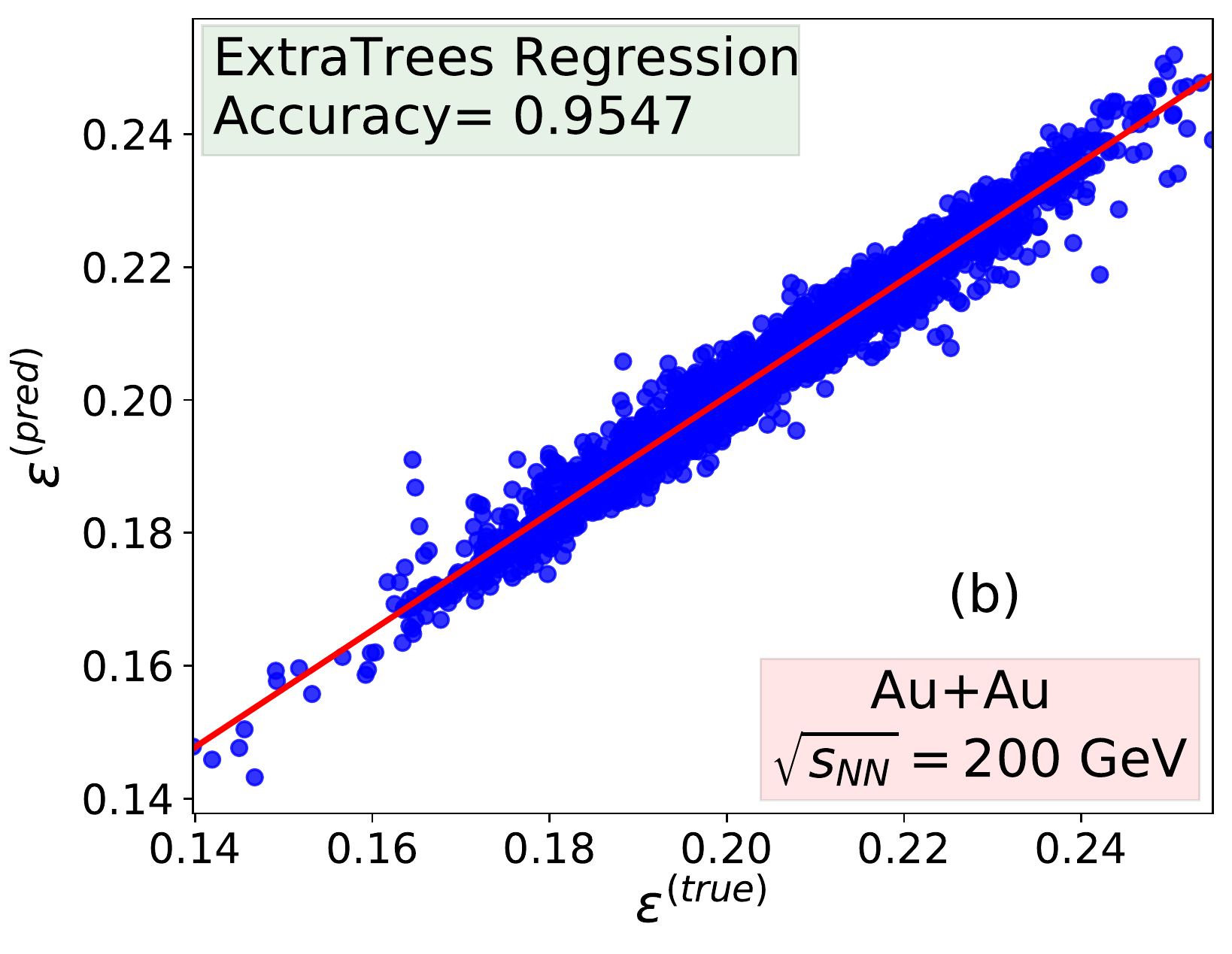}
		\caption{Eccentricity prediction using kNN(a), and ET(b) model with their accuracy score $97.84\%$, and $95.47\%$ for events of Au+Au system at a collision energy $200$ GeV. These plots are obtained for a random train and test set split of input events.}
		\label{fig:eps_pred}
	\end{figure}
	As discussed earlier, the eccentricity is one of the key parameters in heavy-ion collisions. It gives information about initial state geometry and also affects the final state particle flows. But like the impact parameter, it is difficult to measure eccentricity directly from the experiment. Here in this study, the models which are used to get impact parameter prediction, are also used in eccentricity prediction. In fact, ET, kNN, RF are the three best-performing algorithms in the case of eccentricity prediction.
	
%
	Fig. \ref{fig:eps_pred} shows the prediction plot of eccentricity using the kNN(a) and ET(b) model. The accuracies obtained are $97.84\%$ and  $95.47\%$ respectively. This is observed for a randomly split train-test dataset of minimum bias Au+Au events. The models are trained using $10000$ randomly selected events and the testing is performed over $2000$ events which are shown in Fig. \ref{fig:eps_pred}.
	  The 10-fold cross-validation score is also closer to the accuracy obtained using the random train-test split dataset, $97.52\%$ for the kNN model and $95.18\%$ for the ET model. The 10-fold CV score of RF model is $91.95\%$. We get accuracy between $87\%$ to $93\%$ when the ML models are trained using the default AMPT model data.
	
	\begin{figure}
		\includegraphics[width=0.6\linewidth] {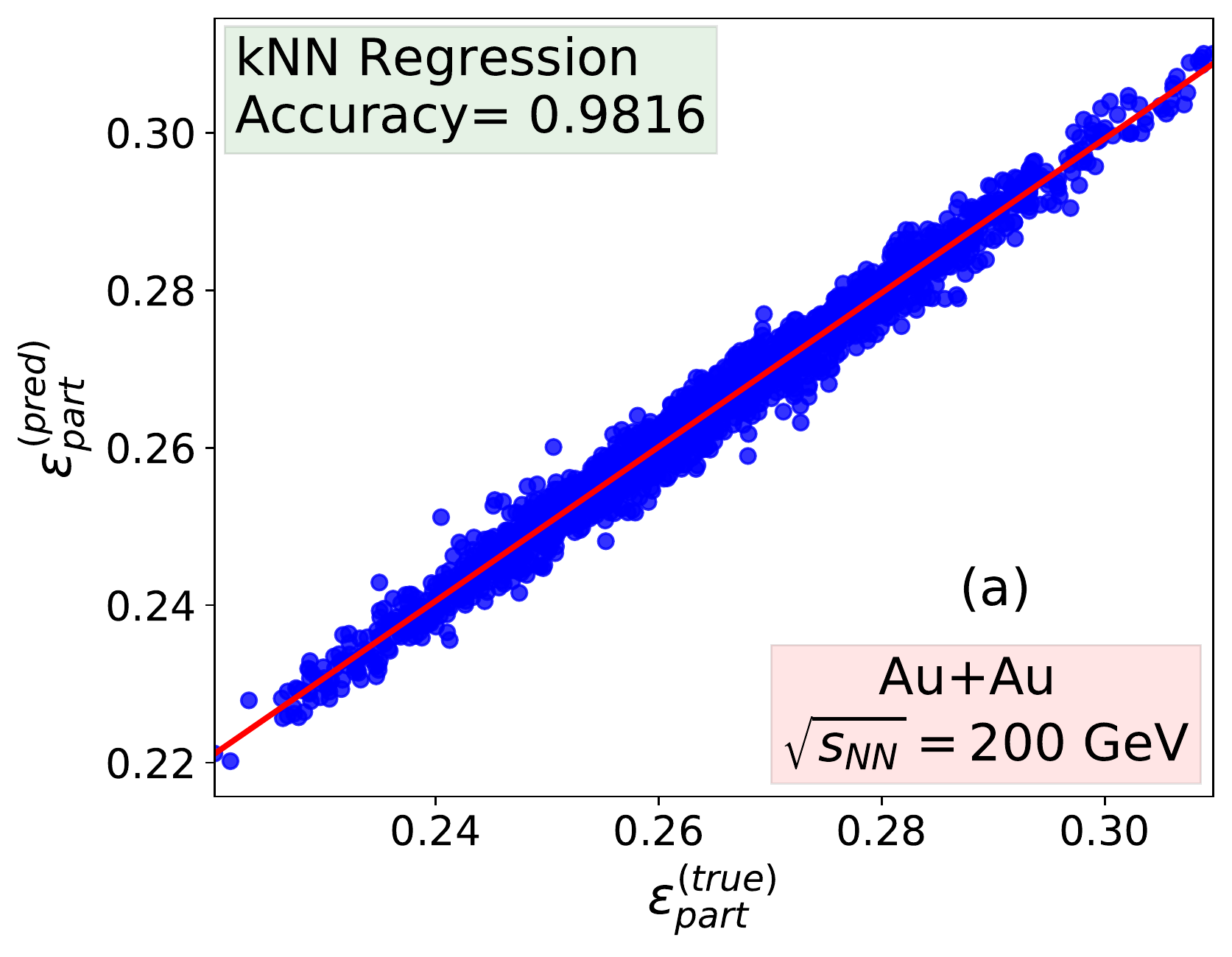}
		\includegraphics[width=0.6\linewidth] {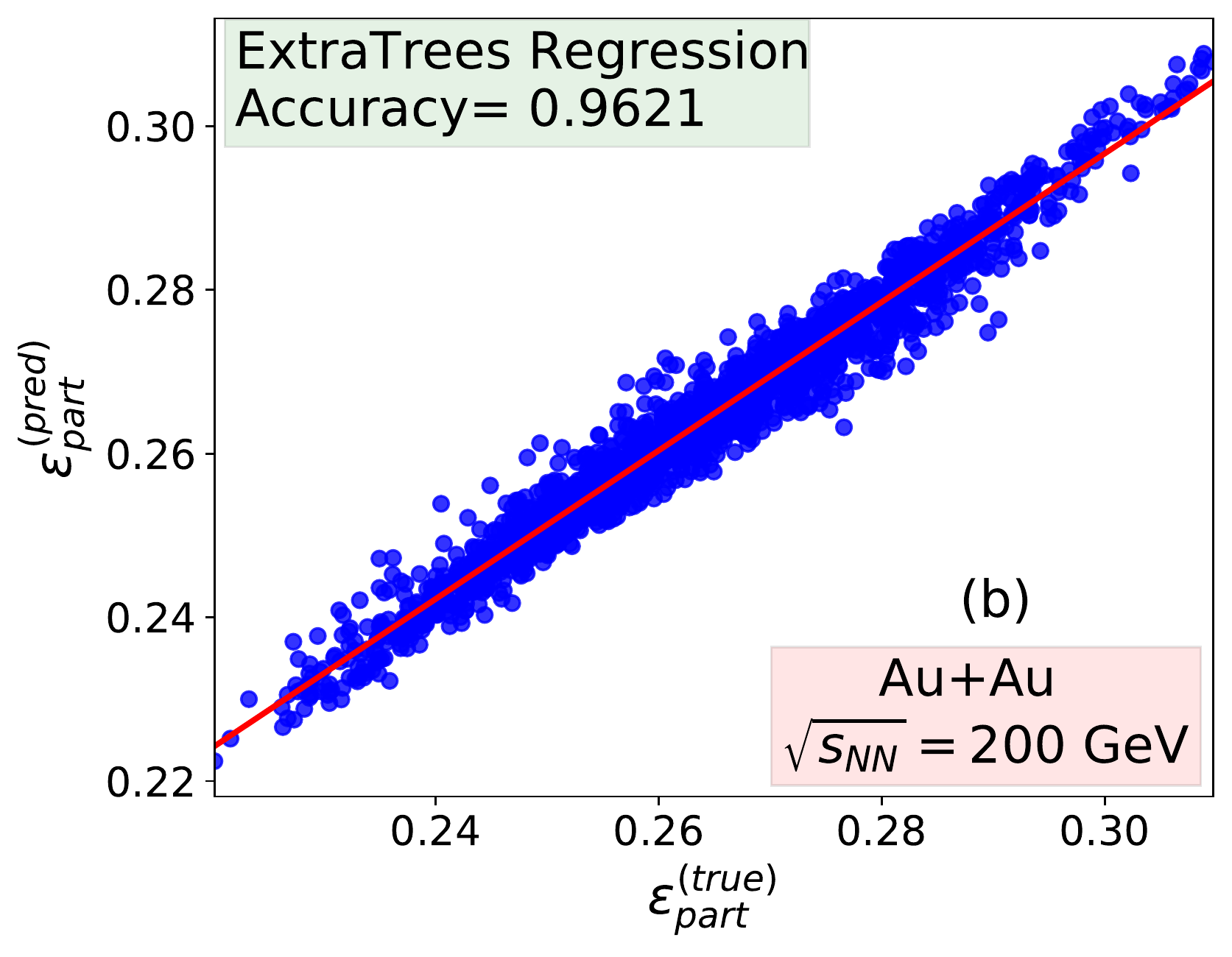}
		\caption{Participant eccentricity prediction using kNN(a), and ET(b) model with their accuracy score $98.16\%$, and $96.21\%$ for events of Au+Au system at collision energy $200$ GeV. These plots are obtained for a random train and test set split of input events.}
		\label{fig:eps_part_pred}
	\end{figure}
	
	Fig. \ref{fig:eps_part_pred} shows the prediction plot of participant eccentricity using kNN(a) and ET(b) model. The accuracies obtained are $98.16\%$ and $96.21\%$ respectively.
	In this case, the 10-fold cross-validation scores are $97.58\%$ and $95.25\%$ for the kNN and ET model respectively and $93.78\%$ for the RF model.
	 In Table-\ref{best_model_eps_3}, a comparison of accuracy for $\epsilon_{3}$(triangularity) is shown. We have used the Eq.\ref{eq:eps} to obtain the $\epsilon_{3}$. Here also kNN, ET and RF models perform better than the other ML models. All of the three have an accuracy of over $90\%$. The LGBM(Light Gradient Boosting Machine) model also has an accuracy of over $88\%$ after a 10-fold cross validation. This is a tree based machine learning model where the tree grows vertically(leaf-wise)\cite{LGBM}. 
	
%
	\begin{table}[ht!]
		\begin{tabular}{|l|c|c|c|}
			\hline
			\bfseries Model & \bfseries $R^{2}$ & \bfseries MAE$\,\,\,$ & \bfseries RMSE$\,\,$  
			\begin{filecontents*}{best_model_eps_3.csv}
Model,MAE,RMSE,R
k-NearestNeighbors Regressor,0.001,0.0013,0.9762
Extra Trees Regressor,0.0013,0.0017,0.9574
Random Forest Regressor,0.0017,0.0023,0.9216
Light Gradient Boosting Machine,0.0022,0.0029,0.8807
Decision Tree Regressor,0.0024,0.0041,0.7581
Gradient Boosting Regressor,0.004,0.005,0.6309
\end{filecontents*}
\csvreader[head to column names]{best_model_eps_3.csv}{}
			{\\\hline\Model & $\,\,$ \R $\,\,$&$\,\,$ \MAE $\,\,$& \RMSE  }\\
			\cline{1-4}
		\end{tabular}
		\caption{10-Fold cross-validation accuracy of ML models for $\epsilon_{3}$ predictions of min. bias Au-Au events at $\sqrt{s}=200$ GeV}
		\label{best_model_eps_3}
	\end{table}

\begin{figure}
		\includegraphics[width=.6\linewidth]{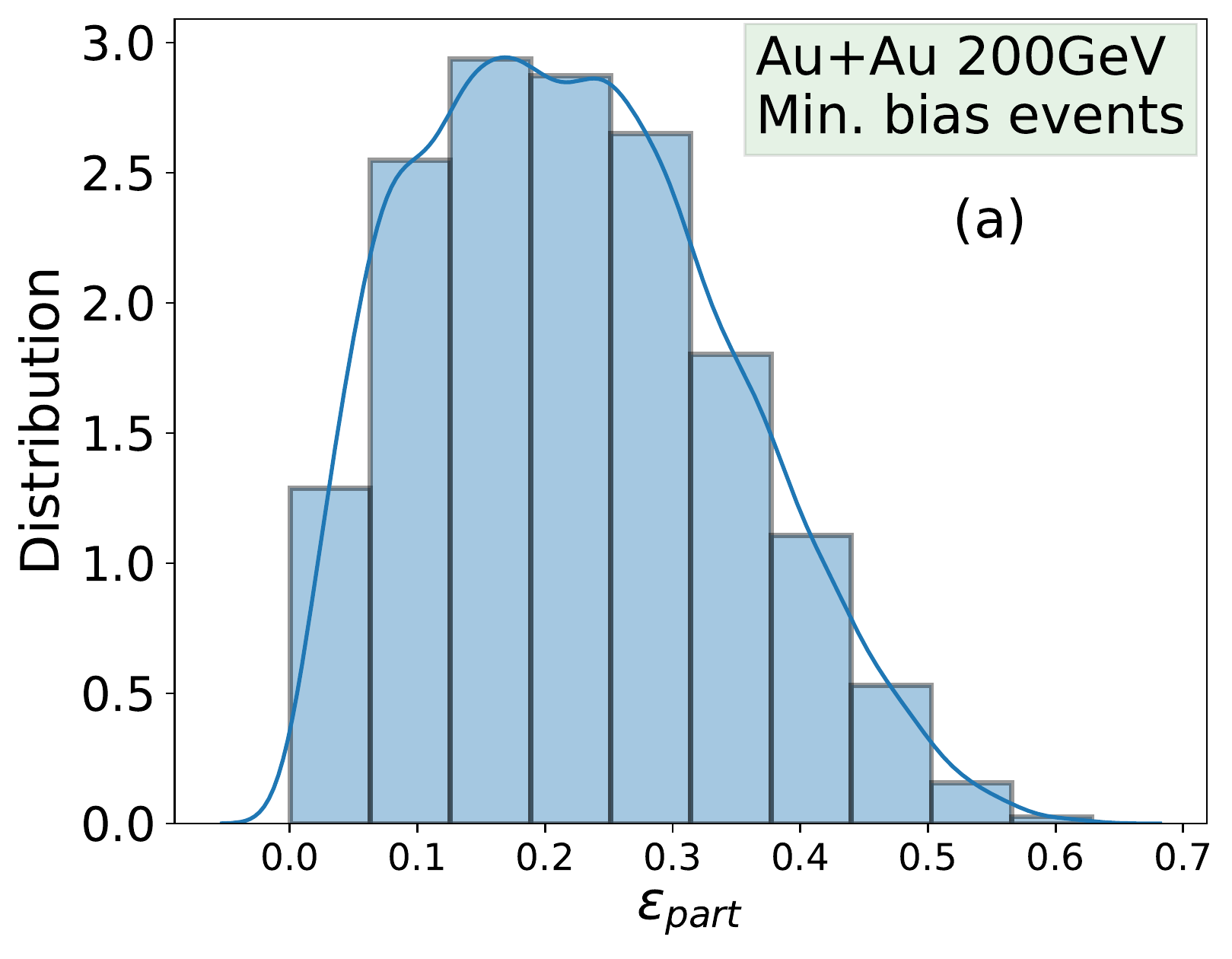}
		\includegraphics[width=.6\linewidth]{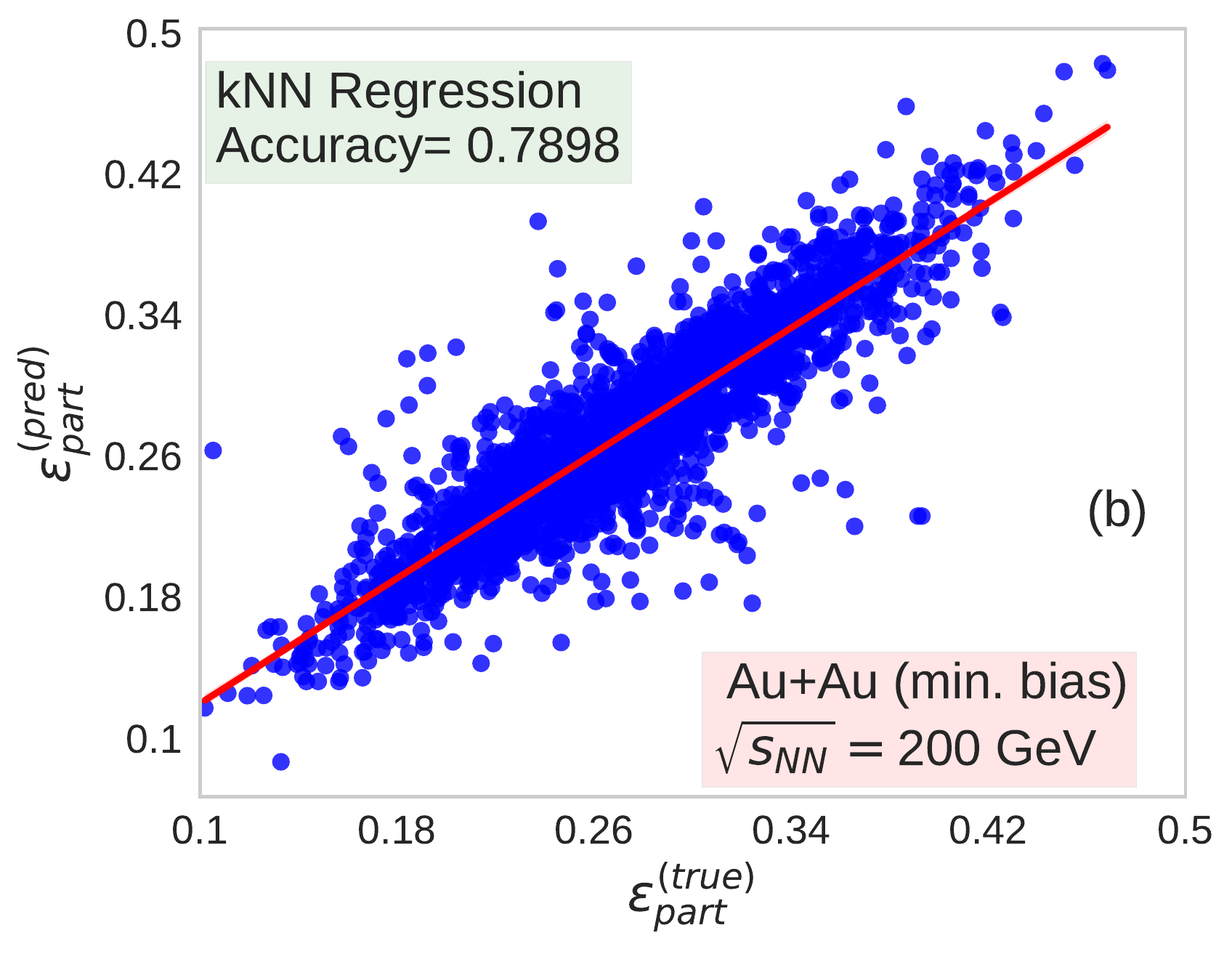}
		\caption{a) Histogram plot of participant eccentricity distribution and b) Prediction plot of $\epsilon_{part}$ for higher $\epsilon_{part}$ range using kNN model of minimum bias Au-Au collision events at $\sqrt{s}=200$ GeV given by the AMPT model}
		\label{fig:eps_dist}
	\end{figure}

	In the eccentricity prediction figures (ref Fig. \ref{fig:eps_part_pred}), a small range of eccentricity ($0.22-0.32$) is taken for the model fitting and predictions. It is specifically the range where the maximum prediction accuracy is obtained for all the models. One of the reasons behind this is that the distribution of eccentricity over the events is not isotropic. In Fig. \ref{fig:eps_dist}(a), the distribution of participant eccentricity is shown. Here the vertical axis represents the normalized number of events, and the horizontal axis gives the eccentricity range. The peak in the distribution is observed for eccentricities between $0.15$ to $0.25$. The distribution is thus skewed, and it means that we have an imbalanced dataset. 
So, the eccentricity of maximum events that are occurring fall in a particular range. Hence the model fits well in this range of eccentricity because of a larger number of fitting points. From the graph, we see that  the range of eccentricity can be increased further from $0.1$ to $0.5$. In Fig. \ref{fig:eps_dist}(b), a prediction plot of participant eccentricity using the kNN model is given for a larger range. Here the events are considered which have eccentricities in the range from $0.1$ to $0.5$. So, the range has now become three times wider than the previous cases. We observe that the points are wider from the center and away from the $45^{0}$ red line compared to the points in Fig. \ref{fig:eps_part_pred}(a). We also see some points which are away and isolated from the bulk distribution. The accuracy is lowered  to $78.98\%$ from its previous value of $98.16\%$. The 10-fold cross-validation score is $76\%$ in this case which is a fair amount of accuracy though it is  much lower compared to the maximum accuracy. This means that the range of accuracy can be fixed according to the requirement of the problem. To accommodate a wider eccentricity range, we have to compensate with accuracy.  We have also applied different ML algorithms to obtain the accuracy at different collision energies from $20$ GeV to $200$ GeV for the impact parameter, eccentricity, and the participant eccentricity predictions. For lower collision energies, the number of events required to train an ML model is higher compared to the number of events required for higher collision energies. This is because high multiplicity events are generated at higher collision energies. Thus, the event-by-event averages become stable. 

\begin{figure}
		\includegraphics[height=0.6\linewidth]{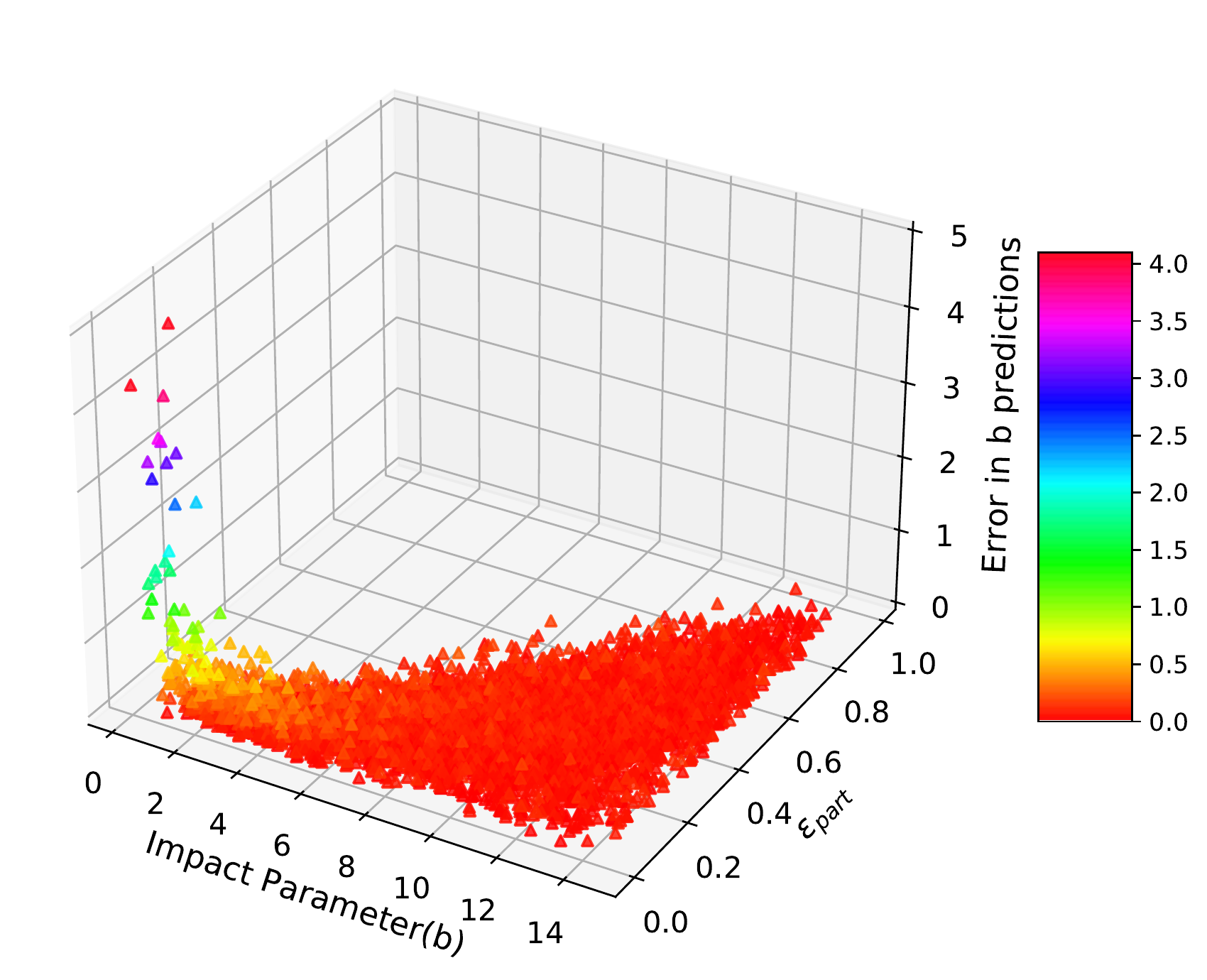}
		\caption{Error in the prediction of impact parameter as a function of impact parameter and eccentricity distribution. This is for $200$ GeV Au-Au collisions and the prediction is obtained using a kNN model}
		\label{fig:ecc_b2}
	\end{figure}

	In Fig. \ref{fig:ecc_b2}, we plot the error in the impact parameter prediction as a function of the impact parameter and the eccentricity distribution. The error here is the relative error(RE) which is given by $RE= \left| \frac{b_{pred}-b_{org}}{b_{org}}\right| $, where $b_{pred}$ and $b_{org}$ are the predicted and original value of impact parameter. We observe that for all eccentricity and impact parameter ranges, the error is low except for the region where the impact parameter is less than $2$fm. In the majority of the distribution, the difference in the prediction and the original impact parameter is less than $0.5$ (shown by the red points) and in some cases, it is less than $1$. But for the lower range of impact parameters($b<2fm$) and eccentricities, we find the difference becomes significantly larger. This is also because of imbalance in the data as discussed. There are comparably large errors for MC-Glauber model predictions of low impact parameters.
	Large discrepencies are also obtained for events of UrQMD and AMPT with higher charge particle when fitted with MC-Glauber model data which is shown in ref.\cite{discr}. In ref. \cite{alicediscr}, large errors are observed for the fitting of Glauber moel data to the ALICE data.
	Our results are similar to the recent results obtained from ML using other models like UrQMD where it was shown that the impact parameters are determined efficiently in all the regions except the very central and the very peripheral regions \cite{tsang2021}. This is adequately reflected in Fig.\ref{fig:ecc_b2}, where a large amount of error is found in the very central region.   
	
\subsection{Results from the different HIC models}	
	
	To check the model dependency, we have used the data from other HIC models and obtained the prediction of the impact parameter. The training of an ML model has been done using the AMPT model data, but the predictions are made for other heavy-ion collision models. The other HIC models used in this study are the VISH2+1 model \cite{vish1} which is a hydrodynamic evolution code and a hybrid model made of VISH2+1 and UrQMD model \cite{iEBE}. In the hybrid model, the UrQMD code is used for later-stage hadronic re-scatterings. The reason for using multiple models is because we want the test set and the training set to come from different models giving the same $p_{T}$ spectra. This would mean that as long as the $p_{T}$ spectra is the same, the ML algorithms will not know which model simulated the test data. As done in the previous cases, here also we have used the transverse momentum spectra of the AMPT model as the features and impact parameters of the corresponding AMPT events as targets for the ML model training. The AMPT events considered here are the minimum bias Au-Au events of $200$ GeV collision energy. The $p_{T}$ spectra of VISH2+1 and hybrid model are obtained at the same collision energy and at specific centralities with impact parameters ranging from $0.1$ fm to $14$ fm. The parameter settings of the VISH2+1 model in this study is similar to the parameters considered in ref. \cite{vish_expt}, with the Glauber initial condition, shear viscosity to entropy density ratio $\eta/s=0.16$ and decoupling temperature $T_{dec}=160 MeV$. The box plot is obtained for s95p-PCE equation of state. For the hybrid model, we set the $\eta/s$ to $0.08$, the equation of state used is s95p-PCE and $T_{dec}=165 MeV$. 
	  We have got $5000$ events each from both of these models at all the impact parameter ranges separately and fitted the average $p_{T}$ spectra with the experimentally obtained $p_{T}$ spectra\cite{spectra0_10, spectra40_80}.  We have considered the $0.15$ to $1.4$ GeV/c $p_{T}$ range to fit with the experimental spectra and also for ML model training. In this range, VISH2+1 data fits well with the experimental data. By doing this we are ensuring that the data is similar to the experiments. In an approximate manner, we are also examining the performance of ML models in case of the use of experimental data as test data for prediction.
	  As we do not have event-by-event experimental data at specific centralities, we have used different HIC models to generate the $p_{T}$ spectra. In this way, we are able to obtain the error distribution of the predictions given by the ML model for a large number of events at specific centralities.
	  
	All the ML models considered in this study, e.g., kNN, RF, ET, and LR, perform reasonably well for impact parameter prediction for an unknown HIC model test data. In Fig. \ref{fig.error_plots}(a) and \ref{fig.error_plots}(b), we show the error plots of impact parameter predictions by the kNN model for VISH2+1 and hybrid UrQMD model respectively. These are relative errors, and the box represents the distribution of errors.  The middle line inside the box represents the median error which is in the middle of the box. The top and bottom lines represent the 25th and 75th percentile of the error distribution. The green point is the mean error. In all the boxes i.e., at all the centralities, the errors are in a normal distribution. This shows a good prediction by the ML model. The end circles are the outliers which are less in number. In both the figures, we find error goes down for the higher impact parameter events. Above the impact parameter of $2$ fm, the prediction errors are very close to zero. Above $b=10$ fm, the errors stayed low continuing the previous trend. The three lines in Fig.\ref{fig.error_plots}(a) show the mean errors in impact parameter prediction for different Equations of State(EoS). The green(dashed), blue(solid), yellow(dotted) line represent the mean errors for s95p-PCE, EOS L and SM-EOS Q equations of state respectively. For all the EoS, the trend of error distribution is similar. In Fig. \ref{fig.error_plots}(a), we see that for $0.1$ and $0.5$ fm impact parameter events, the relative prediction errors are more than three times and in Fig. \ref{fig.error_plots}(b), it is more than five times compared to the real impact parameters. 
	We have considered Au-Au collision events where most central collisions of $0-5\%$ centrality are comprised of events of impact parameter range $0$ to $3.31$ fm \cite{branges}. As has been discussed, in the case of experiments, the centrality is found out using the Glauber model. Thus, it is difficult to assign a specific impact parameter value especially in the case of the most central events. We have seen the same nature in error prediction in Fig. \ref{fig:ecc_b2} while working with only the AMPT events. In that case, training of the ML model and the testing are performed with the AMPT events. This is due to the imbalance in the dataset that we are using for the ML model training. Although we get similar nature of error distribution in all the cases, we used the Glauber initial conditions for the hydro model input. The initial condition from the Color Glass Condensate model can give different $p_{T}$ spectra. To check whether the ML models are effective in this scenario, the parameters of the hydro model should be adjusted such that the $p_{T}$ spectra obtained match well with the experimental $p_{T}$ spectra.
	
\subsection{Results from rebalancing the data set}

	\begin{figure}
		\includegraphics[width=0.6\linewidth]{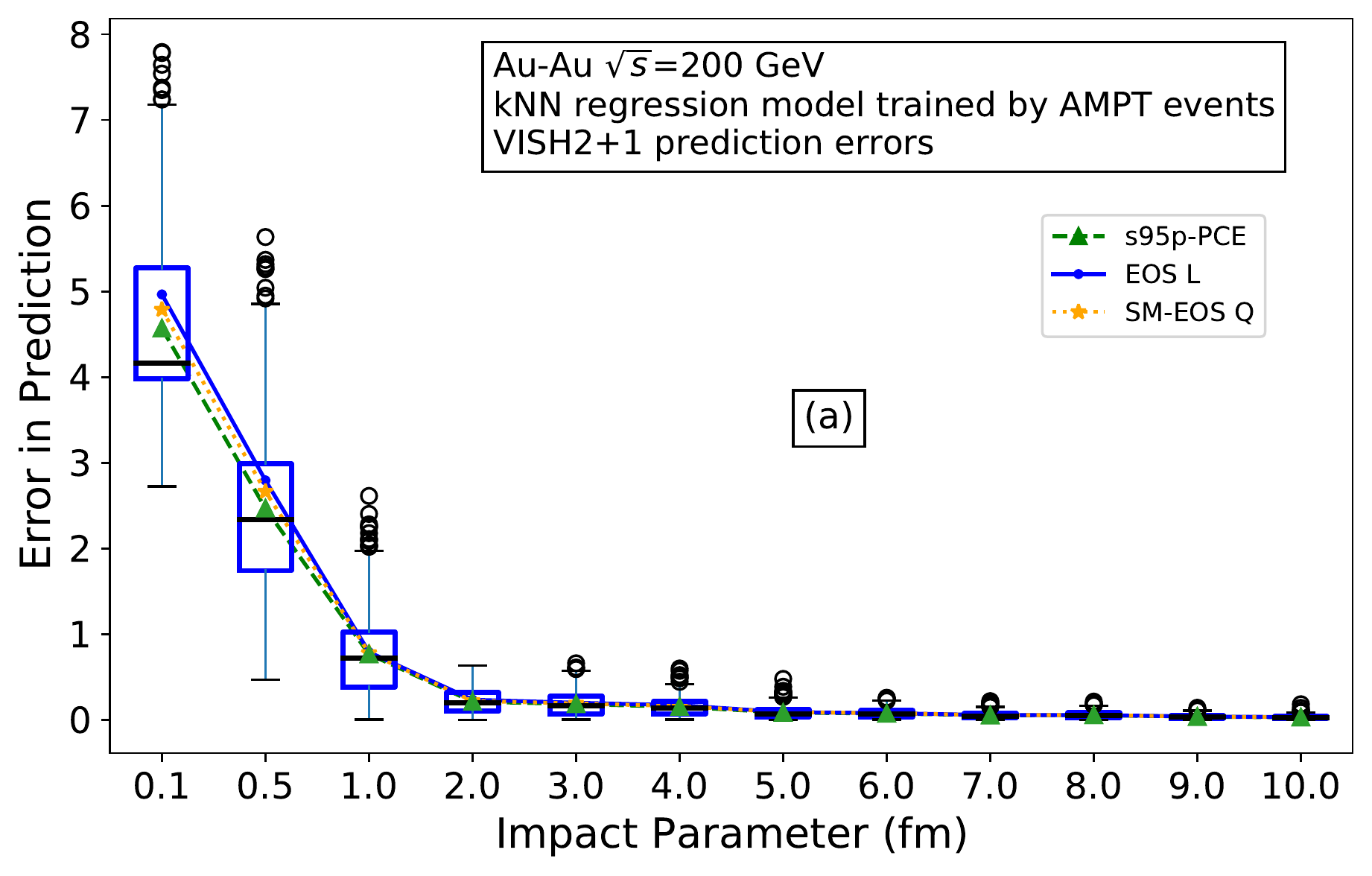}
		\includegraphics[width=0.6\linewidth]{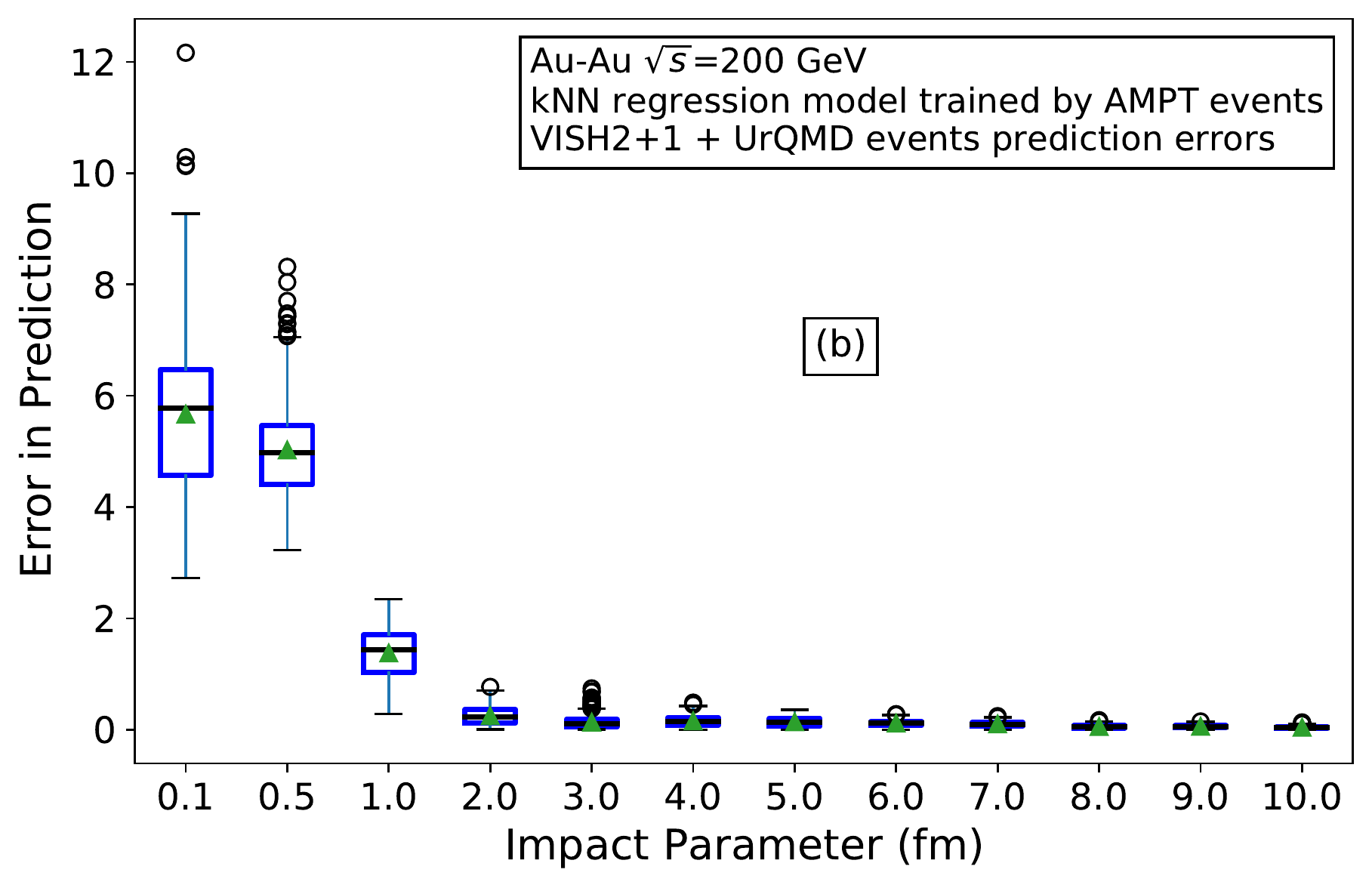}
		\caption{Error plot of impact parameter predictions by kNN model of different centrality events of a) VISH2+1 and b) UrQMD simulations. The 3 lines in (a) show the mean errors in impact parameter prediction for different EoS}
		\label{fig.error_plots}
	\end{figure}

\begin{figure}
		\includegraphics[width=0.6\linewidth]{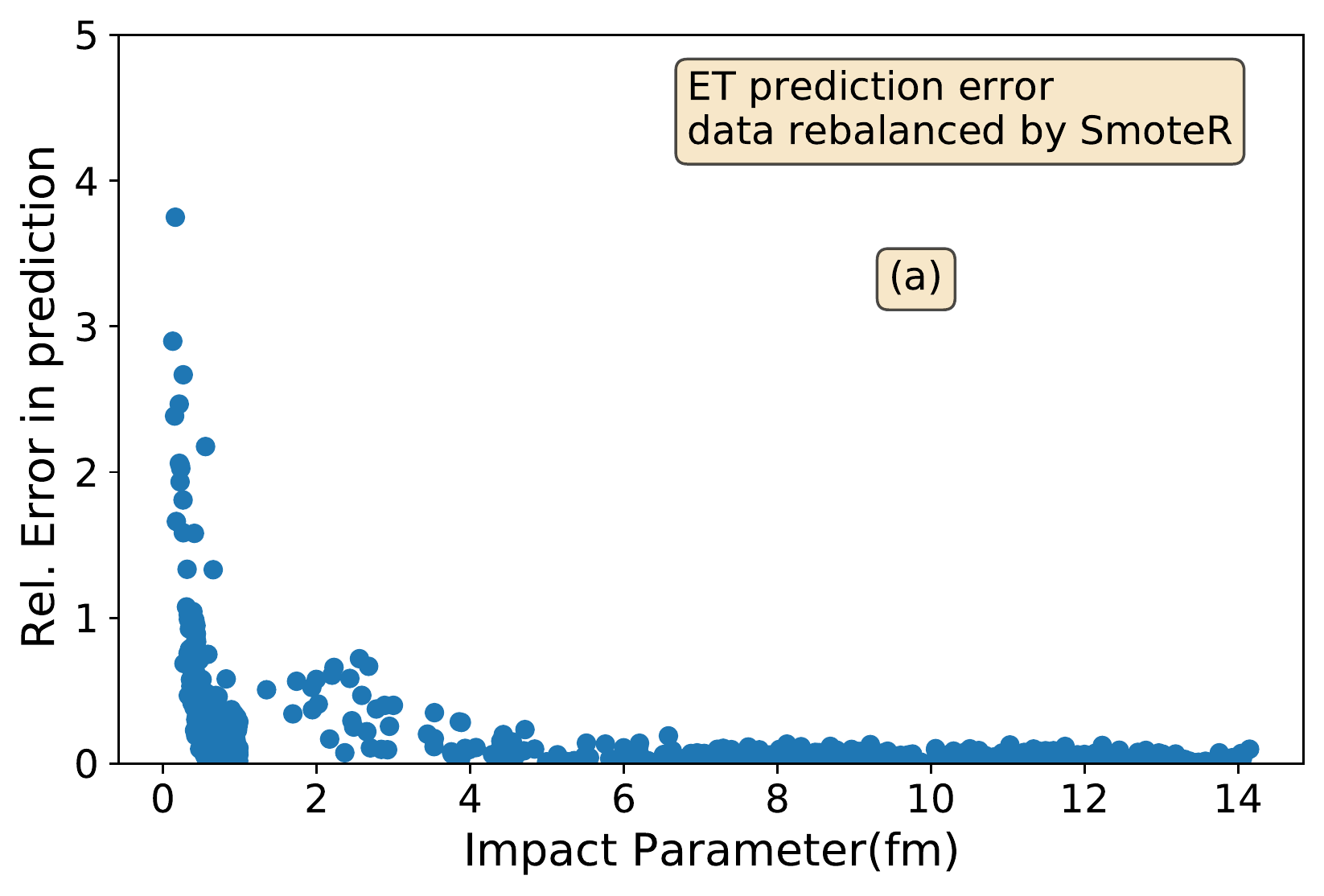}
		\includegraphics[width=0.6\linewidth]{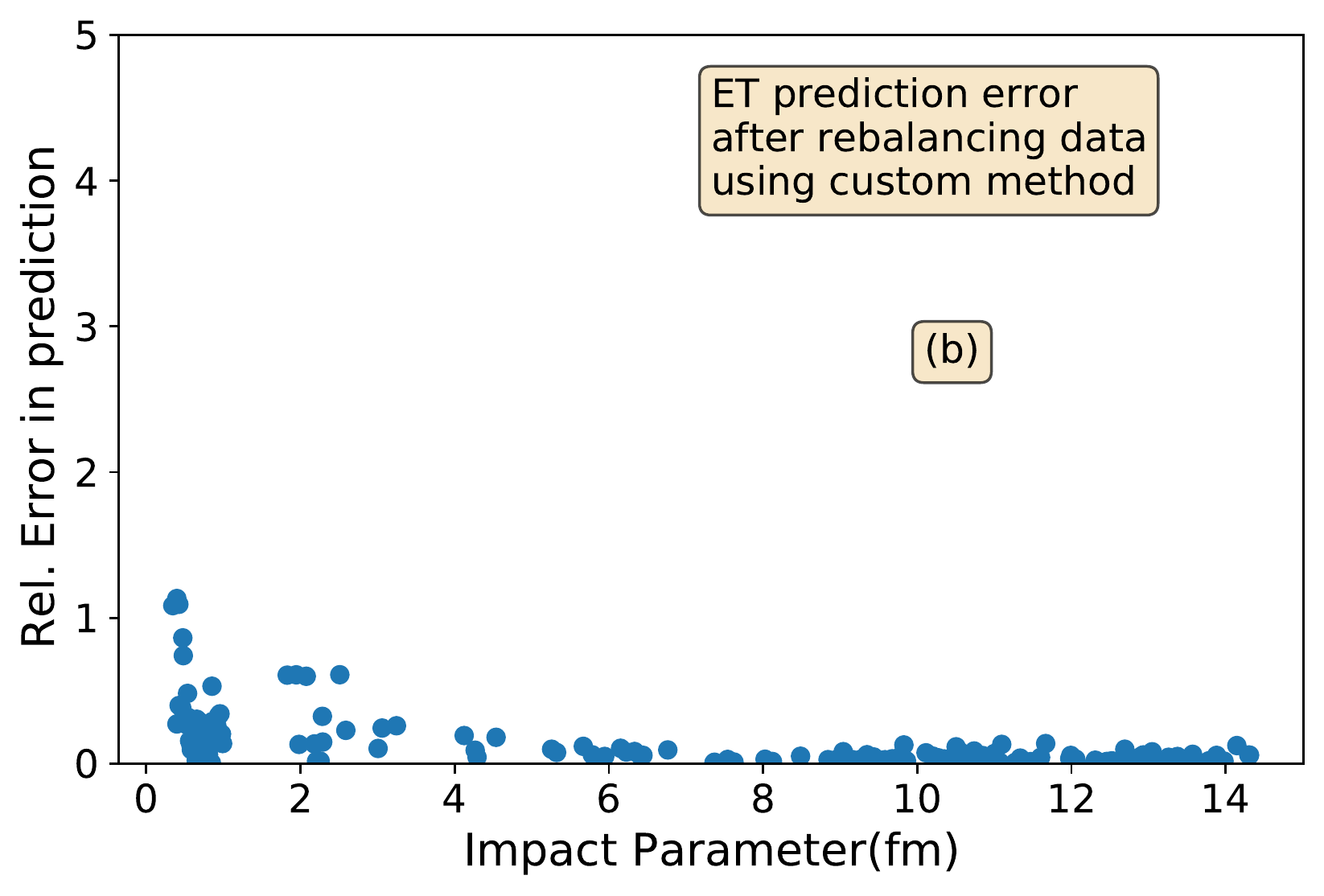}
		\caption{Error distribution of ET model of impact parameter predictions of Au-Au collisions at $\sqrt{s}= 200$ GeV. The training set is re-balanced using a) SmoteR method, and b) a custom method of giving weights to the input data}
		\label{fig.rebalance}
	\end{figure}

	 A large error [Fig. \ref{fig:ecc_b2}, \ref{fig.error_plots}] is observed in the prediction in the lower impact parameter range due to the imbalance in the impact parameter distribution in the training set. We overcome this through a custom sample weighing method as mentioned in Section III (B). As mentioned previously in Section III B, initially, we have used standard packages to rebalance the data.  The results are shown in Fig.\ref{fig.rebalance}(a) for one of the methods. The others also give similar results.    
 Although the error comes down in the lower impact parameter region compared to the errors obtained in Fig.\ref{fig:ecc_b2}, still we get enough errors that would give a wrong estimate for the low impact parameter events.  Finally, we give the results of our custom rebalancing method which has been described previously in detail in Section III B in Fig. \ref{fig.rebalance}(b). 
	 With our custom method we were able to minimize the error to less than 1 as shown in Fig. \ref{fig.rebalance}(b). This error is acceptable in this impact parameter range as the prediction made in this range will always fall in the most central collision category ($0-5\%$) for the Au-Au collisions. 
	
\begin{figure}
		\includegraphics[width=0.48\linewidth]{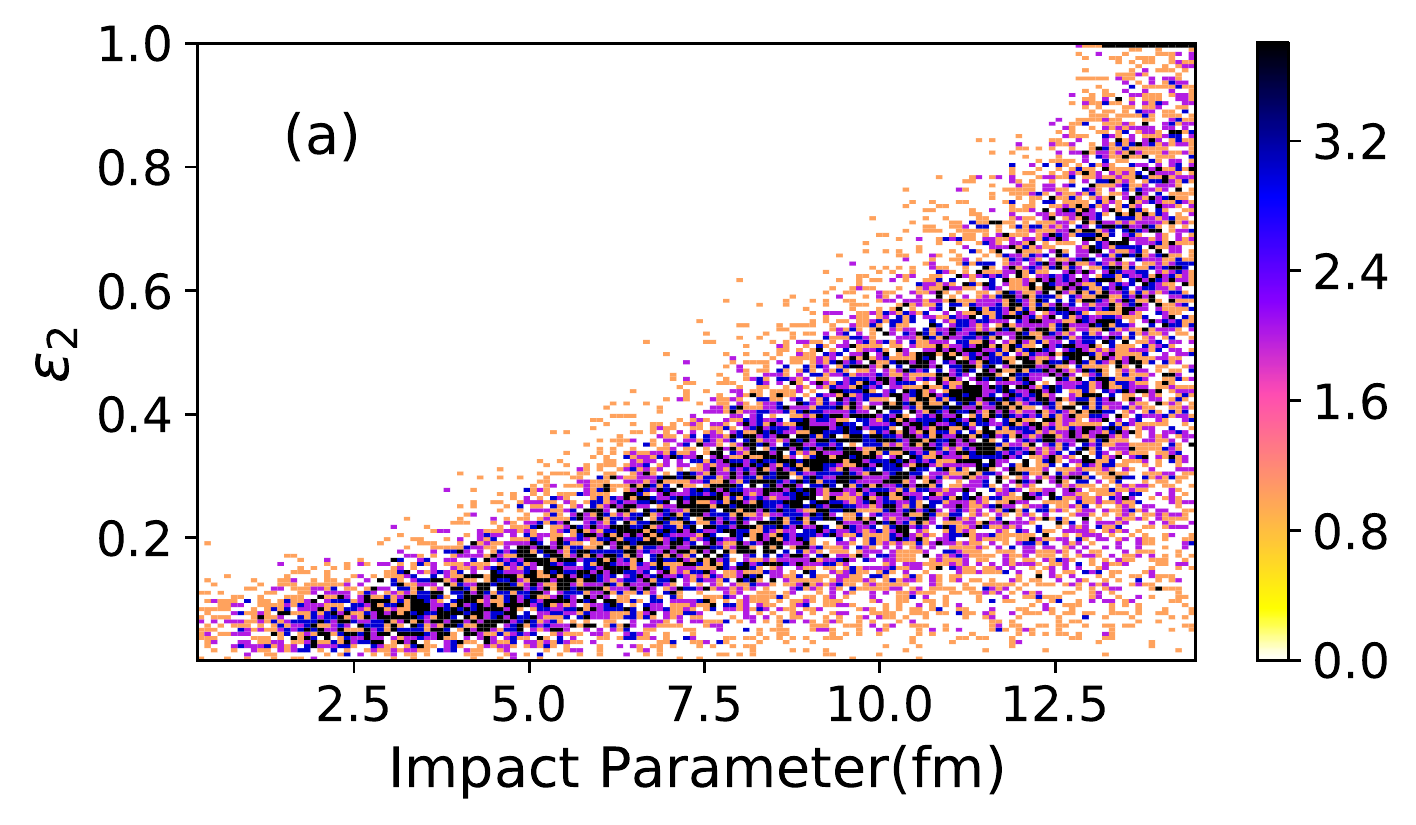} 
		\includegraphics[width=0.48\linewidth]{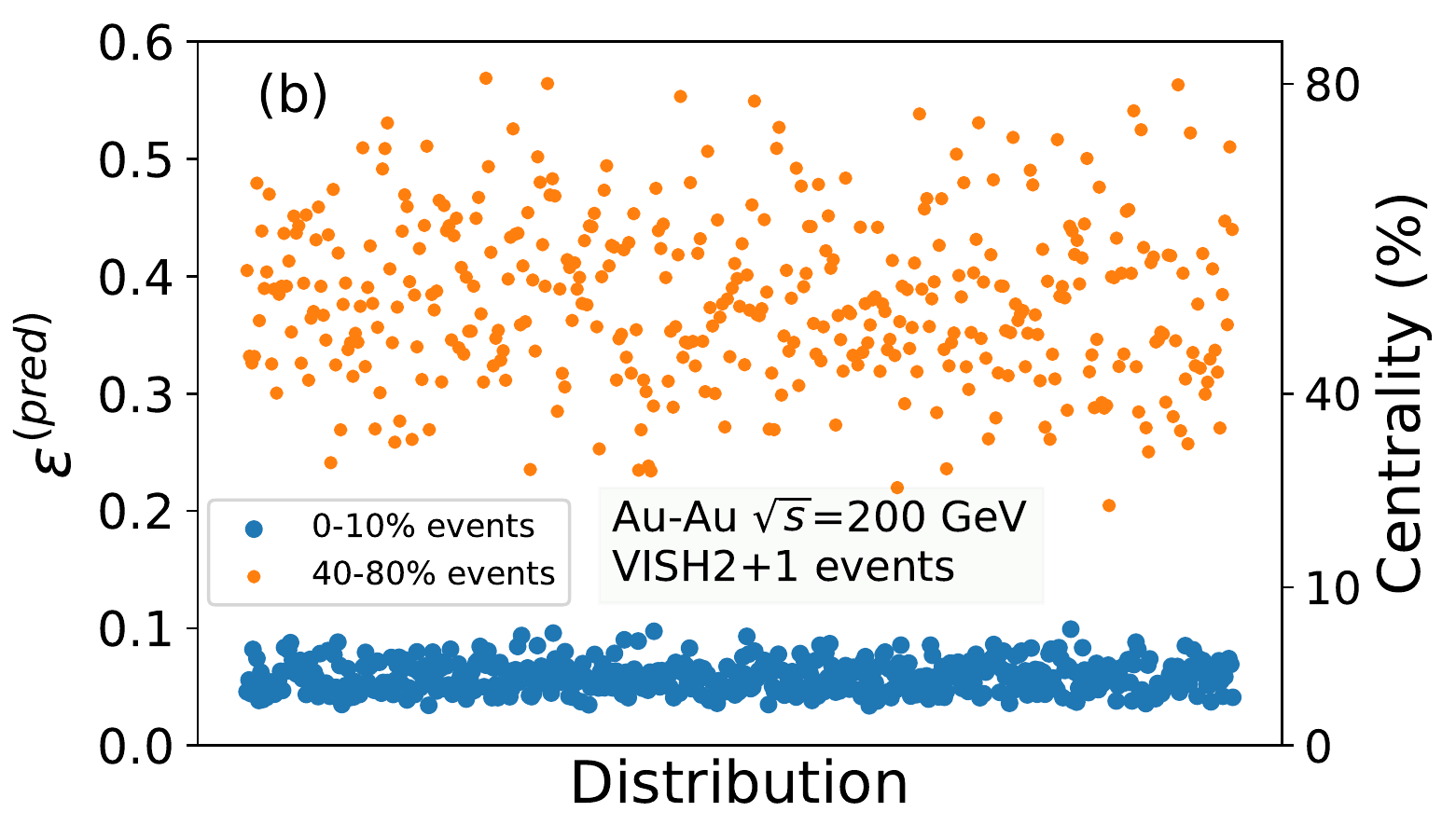}
		\caption{a) Distribution of eccentricity with impact parameter of min. bias Au-Au collision events at $\sqrt{s}= 200$ GeV, b) Distribution of eccentricity predictions by kNN model of $0-10\%$ and $40-80\%$ centrality events of Au-Au collisions at $\sqrt{s}= 200$ GeV from the VISH2+1 model.}
		\label{fig.ecc_vish}
	\end{figure}

	 It is also interesting to see how the AMPT trained models predict eccentricities when they are introduced to other HIC model data. In Fig. \ref{fig.ecc_vish}(a), we show the distribution of eccentricity with the centrality of $200$ GeV AMPT collision events. Although the color plot suggests that there is a linear relationship between the average eccentricity and impact parameter of collision events. We see that the range of eccentricity is lower for lower impact parameter values. As we go for higher impact parameter events, the range of eccentricity becomes larger.  A similar observation has been shown previously in ref. \cite{ecc_b_ref}. In Fig. \ref{fig.ecc_vish}(b), we show the distribution of eccentricity predictions of VISH2+1 events for two centrality range. Here also the ML model is trained using minimum bias AMPT events. The orange dots are the prediction events of $40-80\%$ centrality and the blue dots are the prediction events of $0-10\%$ centrality. In the case of $0-10\%$ centrality range, we get eccentricity distribution in $0-0.15$ range which is also in the range of original distribution shown in Fig. \ref{fig.ecc_vish}(a). We get a larger range in  eccentricity values in the higher impact parameter range, the prediction also gives us the same, are represented by the orange dots. Although this shows the model independence characteristics of the ML models, but it is only examined for the Glauber initial conditions of VISH2+1 model. We have not used the Color Glass Condensate initial conditions as it is known that it gives a larger anisotropy but it would be interesting to see how the model would perform in that case. We plan to look at these in a later work.

\begin{figure}
		\includegraphics[width=0.6\linewidth]{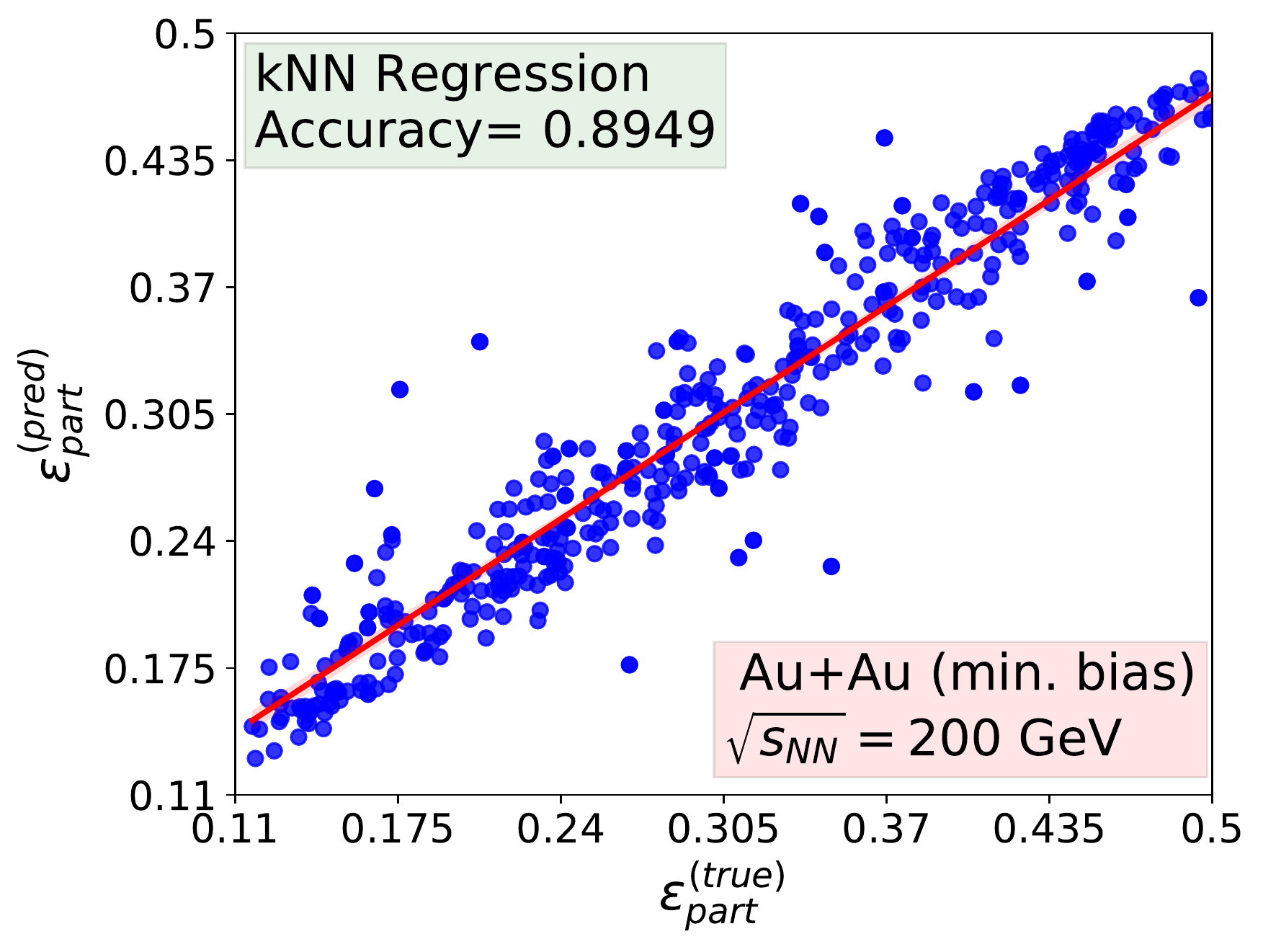}
		\caption{Participant eccentricity predictions of Au-Au collision events at $\sqrt{s}= 200$ GeV after rebalancing the data using the custom method}
		\label{fig.ecc_rebalance}
	\end{figure}

	We have shown an imbalance in the eccentricity distribution in Fig. \ref{fig:eps_dist}(a). Due to this imbalance in the distribution, we get an accuracy of more than $95\%$ in eccentricity prediction only when we consider a small range. For bigger range, the CV accuracy dropped down to $76\%$. A good amount of accuracy can also be achieved for a higher range of eccentricity distribution if the data is rebalanced in a suitable format. We have tried a similar rebalancing technique as is done for the impact parameter prediction. We took the same number of events from each of the distribution bins and trained the model. The prediction plot is shown in Fig. \ref{fig.ecc_rebalance}. We observe that the event points are much more closer to the optimum accuracy line (red line) compared to Fig. \ref{fig:eps_dist}(b) which also have the same range of eccentricity. The accuracy obtained in this case is $89.49\%$ with a cross-validation score of $91\%$. So, using these data rebalancing techniques, one can improve the performance of these ML models for the prediction of the impact parameter as well as the eccentricities.

	\section{Conclusions}
	We have trained different machine learning models to predict various initial stage parameters of a heavy-ion collision system using the AMPT model. We have used the $p_{T}$ spectra for training and testing of the ML models. We have chosen these spectra as it is one of the direct observables in heavy-ion collision experiments. We have observed their learning processes and made changes to the hyperparameters to get an optimum accuracy in the predictions. Out of the various models tested, we have chosen four models kNN, RF, ET and LR for the prediction of the impact parameter. All the models performed well in the impact parameter prediction. Most of the algorithms have shown an accuracy of more than $90\%$ in the prediction of the impact parameter. In the case of the eccentricity, and the participant eccentricity prediction, three models i.e., the kNN, ET, and RF have performed exceptionally well and has given an accuracy of more than $90\%$ after a 10-fold cross-validation. These three models along with the Decision tree and the Light gradient boosting machine have a 10-fold cross-validation score of more than $75\%$ in almost all the cases. 
There is a range of eccentricity ($0.2-0.32$) where the optimum accuracy is obtained for the eccentricity predictions. A greater range of eccentricity ($0.1-0.5$) has also been taken into consideration. We find that the choice of the range in eccentricity affects the prediction accuracy of the eccentricity due to the imbalance in the training data distribution. 
	
	We have also performed an analysis of how the model would possibly perform in predicting the centrality class using experimental data as test data. We have considered two heavy-ion collision models, a viscous hydrodynamic model (VISH2+1) and a hybrid model (Hydro+UrQMD) which are different from the AMPT model that is used for training the machine learning models. The ML model predictions of impact parameters are obtained for the events of the VISH2+1 model and the hybrid model. The hydro and hybrid model events considered for testing are taken at specific impact parameter ranges from $0.5$ fm to $14$ fm. The ML models (kNN results shown specifically) predicted the centrality classes of these events meticulously well. Although in both cases, we have obtained higher errors for the $0.5$ fm events, the errors are very small at other impact parameters. The reason behind this is the lack of balance in the data set. When the data set is normalized, it is found that the peak of the distribution is not at the center. This indicates that the distribution of impact parameter and eccentricity over events are not isotropic.  
	
To minimize these errors, we have used various sampling methods. Though there are several standard packages that help to rebalance the data, we finally see that the accuracy is improved in the lower impact parameter region if we assign different weights to the data at different impact parameters. For the Extratrees model, rarer events are given four times the weightage as the weightage given to impact parameters with a large number of events. This has helped improving the accuracy in the lower impact parameter range. Our rebalancing technique resulted in a cross-validation accuracy of more than $90\%$ for a higher range of eccentricity distribution. This meant an overall improvement from $75\%$ accuracy before the rebalancing to an accuracy of $90\%$ after the rebalancing. Our study therefore shows a rebalanced data set will be useful in making accurate prediction close to the head on collisions. 
	
Finally in conclusion, we have shown that it is possible to use the $p_{T}$ spectra only to make accurate predictions of the initial parameters such as the impact parameter, the eccentricity and the participant eccentricity using the ML algorithms. Even though the algorithm is trained by a single model, it can make accurate predictions from the data generated by other models as long as all the models are able to generate the experimental data accurately. This means that any of the models may be used to train the data set. We have also found that the inaccuracies in the prediction are due to the imbalance in the data set. Proper rebalancing techniques can be used to rebalance the data set and this can be used to predict more accurate results in the low impact parameter regime.

	
	\begin{center}
		Acknowledgments
	\end{center}  
	For computational infrastructure, we acknowledge the Center for Modeling, Simulation and Design (CMSD) at the University of Hyderabad, where part of the simulations was carried out. A.S is supported by INSPIRE Fellowship of the Department of Science and Technology (DST) Govt. of India, through Grant no: IF170627.

\end{document}